\documentclass[12pt,preprint]{aastex}

\usepackage{epsfig}

\shorttitle{Electron temperatures and densities of PNe}
\shortauthors{Y. Zhang et al.}

\begin{document}

%% LaTeX will automatically break titles if they run longer than
%% one line. However, you may use \\ to force a line break if
%% you desire.

\title{Electron temperatures and densities of planetary nebulae determined from
the nebular hydrogen recombination spectrum and temperature and density
variations}

%% Use \author, \affil, and the \and command to format
%% author and affiliation information.
%% Note that \email has replaced the old \authoremail command
%% from AASTeX v4.0. You can use \email to mark an email address
%% anywhere in the paper, not just in the front matter.
%% As in the title, you can use \\ to force line breaks.

\author{Y. Zhang}  
\affil{Department of Astronomy, Peking University, Beijing 100871,
          China}
\affil{National Astronomical Observatories, Chinese Academy of Sciences,
            Beijing 100012,  China}
\email{zhangy@bac.pku.edu.cn}

\author{X.-W. Liu}
\affil{Department of Astronomy, Peking University, Beijing 100871,
          China}

\author{R. Wesson}
\affil{Department of Physics and Astronomy, University College London,
      Gower Street, London WC1E 6BT, UK}

\author{P. J. Storey}
\affil{Department of Physics and Astronomy, University College London,
      Gower Street, London WC1E 6BT, UK}

\author{Y. Liu}
\affil{Department of Astronomy, Peking University, Beijing 100871,
          China}
\affil{Department of physics, Guangzhou University,
       Guangyuan Zhong Lu 248, Guangzhou 510405, China}

\and
\author{I. J. Danziger}
\affil{Osservatorio Astronomico di Trieste, Via G. B. Tiepolo 11, 1-34131
      Trieste, Italy}

%% Notice that each of these authors has alternate affiliations, which
%% are identified by the \altaffilmark after each name.  Specify alternate
%% affiliation information with \altaffiltext, with one command per each
%% affiliation.

%\altaffiltext{1}{present address: Department of Astronomy, Peking university,
%                Beijing 100871, P. R. China}

%% Mark off your abstract in the ``abstract'' environment. In the manuscript
%% style, abstract will output a Received/Accepted line after the
%% title and affiliation information. No date will appear since the author
%% does not have this information. The dates will be filled in by the
%% editorial office after submission.

\begin{abstract}
A method is presented to derive electron temperatures and densities
of planetary nebulae (PNe) simultaneously, using the observed hydrogen
recombination spectrum, which includes continuum and line emission.
By matching theoretical spectra to observed spectra around the Balmer jump
at about 3646~{\AA}, we 
determine electron temperatures and densities for 48 Galactic PNe.
The electron temperatures based on this method [hereafter $T_{\rm e}$(Bal)]
are found to be systematically lower than those derived from [\ion{O}{3}] 
$\lambda4959/\lambda4363$ and
[\ion{O}{3}] $(88\mu{\rm m}+52\mu{\rm m})/\lambda4959$ ratios
[hereafter, $T_{\rm e}$([\ion{O}{3}]$_{\rm na}$) and $T_{\rm e}$([\ion{O}{3}]$_{\rm fn}$)].
And the electron densities based on this method
are found to be systematically higher than those derived from [\ion{O}{2}]
 $\lambda3729/\lambda3726$, [\ion{S}{2}] $\lambda6731/\lambda6716$, 
[\ion{Cl}{3}] $\lambda5537/\lambda5517$, [\ion{Ar}{4}] 
$\lambda4740/\lambda4711$ and
[\ion{O}{3}] $88\mu{\rm m}/52\mu{\rm m}$ ratios. These results suggest that
temperature and density fluctuations are generally present within nebulae.
The comparison of  $T_{\rm e}$([\ion{O}{3}]$_{\rm na}$) and $T_{\rm e}$(Bal)
suggests that the fractional mean-square temperature variation ($t^2$) has
a representative value of 0.031. A majority of temperatures derived
from the [\ion{O}{3}] $(88\mu{\rm m}+52\mu{\rm m})/\lambda4959$ ratio are found to be higher
than $T_{\rm e}$([\ion{O}{3}]$_{\rm na}$), which is attributed to the existence
of dense clumps in nebulae -- those [\ion{O}{3}] IR fine-structure lines
are suppressed by collisional de-excitation in the clumps. By comparing $T_{\rm e}$([\ion{O}{3}]$_{\rm fn}$), 
$T_{\rm e}$([\ion{O}{3}]$_{\rm na}$) and $T_{\rm e}$(Bal) and 
assuming a simple two-density-component model, we find that the filling factor
of dense clumps has a representative value of 7$\times10^{-5}$.
The discrepancies between $T_{\rm e}$([\ion{O}{3}]$_{\rm na}$) and 
$T_{\rm e}$(Bal) are found to be anti-correlated with electron densities
derived from various density indicators; high-density nebulae have the smallest
temperature discrepancies. This suggests that temperature discrepancy is
related to nebular evolution. In addition, 
He/H abundances of PNe are found to be positively correlated with the difference
between $T_{\rm e}$([\ion{O}{3}]$_{\rm na}$) and $T_{\rm e}$(Bal),
suggesting that He/H abundances might have been overestimated
generally because of the possible existence of H-deficient knots.
Electron temperatures and densities deduced from spectra around the Paschen 
jump regions at 8250~{\AA} are also obtained for four PNe, NGC~7027, NGC~6153, M~1-42 and
NGC~7009. Electron densities derived from spectra around the
Paschen jump regions are in good agreement with the corresponding values derived
from spectra around the Balmer jump, whereas temperatures deduced 
from the spectra around the Paschen jump are found to be lower than
the corresponding values derived from spectra around the Balmer jump
for all the four cases. The reason remains unclear.

\end{abstract}

%% Keywords should appear after the \end{abstract} command. The uncommented
%% example has been keyed in ApJ style. See the instructions to authors
%% for the journal to which you are submitting your paper to determine
%% what keyword punctuation is appropriate.


%% From the front matter, we move on to the body of the paper.
%% In the first two sections, notice the use of the natbib \citep
%% and \citet commands to identify citations.  The citations are
%% tied to the reference list via symbolic KEYs. The KEY corresponds
%% to the KEY in the \bibitem in the reference list below. We have
%% chosen the first three characters of the first author's name plus
%% the last two numeral of the year of publication as our KEY for
%% each reference.

\section{Introduction}

Accurate determinations of metal abundances in planetary nebulae (PNe)
are important for understand the chemical evolution of galaxies and the
nucleosynthesis and mixing processes in low- and intermediate-mass
stars. A long-standing problem in nebular abundance
studies has been that the heavy-element abundances derived from 
optical recombination lines (ORLs) are
systematically higher than those derived from 
collisionally excited lines (CELs; see Liu 2001, 2003 for recent reviews). 
In extreme cases, the discrepancies exceed a factor of 10. 
Given that CEL
abundances are sensitive to electron temperature and density, knowledge of
temperature and density variations is an essential ingredient in
understanding the abundance discrepancy problem. Information on temperature and density 
variations can be obtained by comparing results derived from plasma 
diagnostics which behave differently as temperature and density vary. Traditionally,
nebular electron temperatures and densities are derived from CEL ratios. However,
for PNe including high-density regions ($N$e$\ga10^6$\,cm$^{-3}$), such as 
IC~4997 \citep{hyung1994}, Mz~3 \citep{zhang02} and M~2-24 \citep{zhang2003}, 
CELs are strongly suppressed by collisional de-excitation, and so depending on
their
critical densities, they may become unusable in probing such
high-density regions. On the other hand, hydrogen recombination lines have
very high critical densities ($\ga10^8$\,cm$^{-3}$). Consequently, the
hydrogen recombination spectrum provides a powerful tool to probe 
high-density regions. Comparison of electron 
temperatures and densities derived from the hydrogen recombination spectrum
(lines plus continuum) with those derived from CELs should enable us to 
quantify temperature and
density variations in nebulae.

By measuring the Balmer discontinuity, \citet{liu93} determined electron 
temperatures in 14 PNe and found that they are systematically lower than 
those derived from the [\ion{O}{3}] nebular-to-auroral forbidden line ratio.
This can be explained by the presence of large temperature
fluctuations since the [\ion{O}{3}] $\lambda4363$ auroral line of higher
excitation energy tends to be emitted
in higher temperature regions than the [\ion{O}{3}] 
$\lambda\lambda4959,5007$ nebular lines of lower excitation energy, as suggested by \citet{peim67}.  
Recent reviews on temperature fluctuations of gaseous nebulae are given by
\citet{peim03} and references therein. However, no photoionization
models yield such large temperature fluctuations suggested by observations.
Moreover, temperature fluctuations are unable to explain the large disparity between
the ORL and CEL abundances observed in some PNe \citep{liua}. Instead, it is
found that the hypothesis that PNe may contain 
clumps of different temperature, density and chemical composition,
embedded in diffuse `normal' material, can explain many of the observed
patterns 
\citep{liubarlow2000}. Peimbert's concept of temperature fluctuations alone
may not be
a realistic description of the physical conditions in a nebula, although it
has proven to be a convenient and useful way to characterize the complexities of
real nebulae. 

It is expected that there are also variations of electron density
inside a PN.  \citet{rubin1989} pointed out that electron densities derived from
various plasma diagnostics should correlate positively with the critical densities
of the diagnostic lines involved if there are large density variations within a nebula.
\citet{liubarlow2001} found that electron densities
derived from the [\ion{O}{3}] $88\mu{\rm m}/52\mu{\rm m}$ ratios
are generally lower than those derived from the optical [\ion{Ar}{4}]
and [\ion{Cl}{3}] doublet ratios, suggesting the presence of high density
regions in PNe, where the 52- and 88-$\mu{\rm m}$ lines are suppressed by
collisional de-excitation due to their relatively low critical densities.
\citet{viegas} showed that the discrepancies between
$T_{\rm e}$([\ion{O}{3}]$_{\rm na}$) and
$T_{\rm e}$(Bal) can be explained by the presence of condensations in 
nebulae and gave an analytical method of estimating the filling
factor of condensations. Our detailed studies of the peculiar PNe Mz~3
and M~2-24 suggest that the presence
of high-density regions in some PNe can significantly affect abundance determinations
\citep{zhang02,zhang2003}. However, 
density inhomogeneities alone cannot explain the large disparity between
the ORL and CEL abundances either, unless the condensations
are also H-deficient \citep{liubarlow2000}.

\citet{liubarlow2000} found that the discrepancy between the ORL and CEL
abundances correlates with the difference between the temperatures derived 
from the [\ion{O}{3}] collisional lines and from the Balmer jump of hydrogen 
recombination
spectrum, thus lending strong support to the argument that some extremely cold
H-deficient knots may co-exist with normal nebular material in many PNe.
However, it is difficult to explain the origin and evolution of such
H-deficient knots with the current theories of element production in
intermediate-mass stars. 

In this paper, we present determinations of electron temperatures and
densities for a large sample of PNe based on the analysis of nebular hydrogen recombination spectrum. Section 2 describes the sample. 
In Section 3,
we describe a new approach to deriving electron temperature and
density simultaneously from the observed nebular hydrogen recombination 
spectrum (lines plus continuum) and present the results. 
In Section 4, we compare results derived from a variety of temperature
and density indicators and discuss the significance of variations in
electron temperature and density in PNe. A discussion of He/H abundance 
determinations is also presented in this Section. In Section 5, we present 
electron temperatures and densities derived from the Paschen discontinuity and decrements. A summary
then follows in Section 6.

\section{The sample}

\subsection{Observations and data reduction}

48 Galactic PNe (37 disc PNe and 11 bulge PNe) are analyzed in the current work.
The spectra were obtained in a series of
observing runs using telescopes at ESO La Silla and La Palma.

14 objects were observed using the ISIS double long-slit spectrograph 
mounted on the 4.2\,m William Herschel Telescope (WHT) at
La Palma Observatory in 1996 July--August and in 1997 August.
A Tek
$1024\times1024$ $24\micron\times24\micron$ charge coupled device (CCD) was used, which yielded a spatial
sampling of 0.3576 arcsec per pixel projected on the sky.
In 1996, a 600\,g\,mm$^{-1}$ and a 316\,g\,mm$^{-1}$
grating were used for the Blue and Red Arm, respectively. In 1997,
they were replaced by a 1200\,g\,mm$^{-1}$ grating.

23 objects were observed with the ESO 1.52\,m telescope using the
Boller \& Chivens (B\&C) long-slit spectrograph. The spectra were
obtained during runs in 1995 July, 1996 July, 1997 February and 2001
June. In 1995 the detector was a Ford $2048\times2048$ 
$15\micron\times15\micron$ 
chip, which was superseded in 1996 by a thinned UV-enhanced Loral
$2048\times2048$ $15\micron\times15\micron$ chip of much improved quantum efficiency.
The B\&C spectrograph has a useful slit length of about 3.5\,arcmin. In
order to reduce the CCD read-out noise, the CCD was binned by a factor
of two along the slit direction, yielding a spatial sampling of 1.63~arcsec
per pixel projected on the sky. A slit width of 2\,arcsec was used
throughout the ESO 1.52\,m runs.

11 objects were observed using the Intermediate Dispersion Spectrograph (IDS) on the 
2.5m Isaac Newton Telescope (INT) at the Observatorio del Roque de los 
Muchachos, on La Palma, Spain in 2001 August. A slit 4\,arcmin long and 
1\,arcsec wide was used for nebular observations. The R1200B grating was used to 
cover wavelengths from 3500--5000\,{\AA} at a spectral resolution of 0.95\,{\AA},
determined from the {\it FWHM} of the calibration arc lines, while the R300V 
grating, together with a GG385 order sorting filter, was 
used to cover wavelengths from 3800--7700\,{\AA} at a resolution of 4.0\,{\AA} {\it FWHM}.

All the spectra were reduced using {\sc MIDAS}\footnote{{\sc MIDAS} is developed and distributed by the European
Southern Observatory.}
or {\sc IRAF}\footnote{{\sc IRAF} is developed and distributed by the National
Optical astronomy Observatory.} following the standard procedure. The spectra were
bias-subtracted, flat-fielded and cosmic-rays removed, and then wavelength
calibrated using exposures of a calibration arc lamp.
The spectra were flux-calibrated by observing spectrophotometric standard stars.
1-D spectra were obtained by integrating along the slit. Using the Galactic 
reddening law of \citet{howarth}, the logarithmic extinction at H$\beta$,
$c=\log I({\rm H}\beta)/F({\rm H}\beta)$, was
derived from the observed Balmer decrements, the H$\alpha$/H$\beta$,
H$\gamma$/H$\beta$ and H$\delta$/H$\beta$ ratios. A mean value
for each nebula was used to deredden the spectra.
The object names and the extinction constants are given in Table~\ref{tbl-1}.

Amongst the PNe in our sample, far-infrared spectra for 21 of them 
have been obtained using
the Long Wavelength Spectrometer (LWS) on board the
Infrared Space Observatory ({\it ISO}). Fluxes of far-IR fine structure
lines derived from these spectrum have been published by
\citet{liubarlow2001} except for three objects,
Hu~1-2, NGC~6210 and NGC~6818. For the latter three PNe, we measured 
far-IR line fluxes using spectra retrieved from the {\it ISO}
data archive.
In order to normalize the $ISO$ line fluxes to H$\beta$,
the absolute ${\rm H}\beta$ fluxes tabulated in \citet{cahn1992}
were used.

\subsection{He$^+$/H$^+$ and He$^{2+}$/H$^+$ abundance ratios}

In order to calculate theoretical nebular spectra, accurate determinations of
the He$^+$/H$^+$ and He$^{2+}$/H$^+$ abundance ratios are necessary 
(see Section~3).
Helium ionic abundances were derived from \ion{He}{1} and \ion{He}{2} recombination lines, using
\begin{equation}
\frac{N({\rm He}^{{\rm i}+})}{N({\rm H}^+)}=\frac{I_{\lambda}}{I_{{\rm H}\beta}}\frac{\lambda}{\lambda_{{\rm H}\beta}}\frac{\alpha_{{\rm H}\beta}}{\alpha_{{\rm He}^{{\rm i}+}_\lambda}},
\end{equation}
which have
only a weak dependence on the adopted electron temperature and density.

He$^+$/H$^+$ abundance ratios were calculated from the strong line \ion{He}{1}$\lambda5876$ under 
the assumption of Case A. The effective recombination coefficient 
$\alpha_{{\rm He}^{+}_{5876}}$ is derived from the formulae presented by \citet{benjamin}. 
He$^{2+}$/H$^+$ abundance ratios were calculated from the 
\ion{He}{2}$\lambda4686$ line, using the effective recombination coefficient
$\alpha_{{\rm He}^{2+}_{4686}}$ of \citet{storey1995}. The results are presented
in Table~\ref{tbl-1}.

\subsection{Temperature and Density determined from far-IR and optical forbidden line ratios}

For the purpose of comparison, various forbidden line ratios are also used
to determine electron temperatures and densities.
For densities, results derived from the following diagnostic ratios were
used,
[\ion{O}{3}] $52~\mu$m/$88~\mu$m, [\ion{O}{2}] $\lambda3729/\lambda3726$,
[\ion{S}{2}] $\lambda6716/\lambda6731$, [\ion{Cl}{3}] $\lambda5517/\lambda5537$
and [\ion{Ar}{4}] $\lambda4711/\lambda4740$.
Electron densities from the [\ion{O}{3}]
52-$\mu$m/88-$\mu$m ratio
are taken from \citet{liubarlow2001} except for NGC~6210 and NGC~6818, for which 
we obtain the values of $\log N_{\rm e}=3.19$ and 2.86 (cm$^{-3}$), respectively, using our own measured line fluxes. Densities deduced from
the optical [\ion{O}{2}] $\lambda3729/\lambda3726$,
[\ion{S}{2}] $\lambda6716/\lambda6731$, [\ion{Cl}{3}] $\lambda5517/\lambda5537$
and [\ion{Ar}{4}] $\lambda4711/\lambda4740$ doublet ratios are taken from 
Wang et al. (in preparation).
A detailed discussion of these density indicators is given in their paper.

Electron temperatures derived from the [\ion{O}{3}]
$(88\mu{\rm m}+52\mu{\rm m})/\lambda4959$ and [\ion{O}{3}]
$\lambda4959/\lambda4363$ ratios together with their uncertainties are 
presented in
Table~\ref{tbl-1}. For the former, values were derived assuming electron densities
derived from the [\ion{O}{3}] 52-$\mu$m/88-$\mu$m line ratio.
For the latter, the average densities derived from
the [\ion{Cl}{3}] $\lambda5517/\lambda5537$
and [\ion{Ar}{4}] $\lambda4711/\lambda4740$ doublet line ratios
were adopted, given that Cl$^+$ and Ar$^{2+}$ have ionization potentials of
23.8~eV and 40.7~eV, respectively, comparable to the value of 35.1~eV
for O$^+$, and therefore anticipating that the [\ion{O}{3}] lines arise from similar ionized
regions as [\ion{Cl}{3}] and [\ion{Ar}{4}] lines. As the
density increases, the resultant temperatures become
increasingly sensitive to the adopted density.
When the density reaches a value higher than $10^6$~cm$^{-3}$, the [\ion{O}{3}]
$(88\mu{\rm m}+52\mu{\rm m})/\lambda4959$ and [\ion{O}{3}]
$\lambda4959/\lambda4363$ ratios are no longer useful temperature
diagnostics.
For example, high density regions have been detected in M~2-24 and IC~4997
\citep{zhang2003,hyung1994}.
For both PNe, temperatures deduced from the
[\ion{O}{3}] $\lambda4959/\lambda4363$ ratio have large uncertainties.

%% In this section, we use  the \subsection command to set off
%% a subsection.  \footnote is used to insert a footnote to the text.

%% Observe the use of the LaTeX \label
%% command after the \subsection to give a symbolic KEY to the
%% subsection for cross-referencing in a \ref command.
%% You can use LaTeX's \ref and \label commands to keep track of
%% cross-references to sections, equations, tables, and figures.
%% That way, if you change the order of any elements, LaTeX will
%% automatically renumber them.

%% This section also includes several of the displayed math environments
%% mentioned in the Author Guide.

\section{Electron temperature and density determined from the
hydrogen recombination spectrum}

The \ion{H}{1} recombination spectrum consists of continuum and line
emission. The continuous spectrum has a jump at the limit of the Balmer
series at 3646\,{\AA}. High-order Balmer lines
($n\rightarrow2$, $n\ga25$) converge at the Balmer discontinuity and
merge with the continuum jump at shorter wavelength.
The size of the Balmer jump is a function of
electron temperature. On the other hand, intensities of high-order Balmer lines 
($n\rightarrow2$, $n\ga10$) relative to H$\beta$ are sensitive to electron 
density.
Therefore, the hydrogen recombination spectrum
near the Balmer jump region provides valuable temperature and density diagnostics.
Our method and results of the electron temperature and density 
determinations based on this technique are presented below.

\subsection{Theoretical spectra}

  Several processes contribute to the continuum emission from a nebula:
free-free and free-bound emission and the
two-photon decay. Apart from hydrogen, the second most abundant element, helium 
also contributes significantly to the first two processes.
In addition, emission from the central star and local dust can in principle
contaminate the observed continuum spectrum. In the following, a detailed
calculation including all the processes is presented.

The free-free (or bremsstrahlung) emission, which is emitted when an free
electron interacts but does not combine with a positively charged ion such as H$^+$, He$^+$ and He$^{2+}$, dominates the 
nebular continuum in the infrared and radio region. The emission coefficient 
of this process is given
by \citet{brown},
\begin{equation}
 j_{\nu,ff}=\frac{1}{4\pi}N_iN_e\frac{32Z^2e^4h}{3m_e^2c^3}(\frac{\pi h\nu_0}{3kT_e})^{1/2}e^{-h\nu/kT_e}g_{ff}~[{\rm erg~cm}^{-3}~{\rm s}^{-1}~{\rm Hz}^{-1}],
\end{equation}
where $N_i$ and $N_e$ are ion and electron densities respectively, $Z$ is the
nuclear charge, $T_e$ is the electron temperature, $h\nu_0$ denotes the
Rydberg energy and $g_{ff}$ is the free-free Gaunt factor. Here $g_{ff}$ is
computed using the method outlined by \citet{hummer1988}. 

In the optical region, 
free-bound emission is the main contributor to the continuum.
According to \citet{brown}, the emission coefficient of continuous emission 
due to recombinations of H$^+$, He$^+$ or He$^{2+}$ with electrons is  
\begin{equation}
 j_{\nu,bf}=\frac{1}{4\pi}N_iN_e\frac{\pi h^4\nu^3}{c^2}(\frac{2}{\pi m_ekT_e})^{3/2}\sum_{n=n_1}^{\infty}\sum_{l=0}^{n-1}n^2a_{nl}(\nu)e^{h(\nu _{nl}-\nu)/kT_e}~[{\rm erg~cm}^{-3}~{\rm s}^{-1}~{\rm Hz}^{-1}],
\end{equation}
where $h\nu _{nl}$ is the ionization potential of the ($n$,$l$) state, $n_1$
is the lowest state that can contribute to the free-bound emission at the 
given frequency $\nu$ and $a_{nl}(\nu)$ is the photoionization cross-section
for the state, which is computed using the method described by 
\citet{storey1991}. 

Another important source of continuous emission is
two-photon decay of the 2$^2S$ level of the hydrogen atom
(two-photon emission from He$^+$ and He$^0$ is
negligible for nebulae of cosmic compositions; Brown $\&$ Mathews, 1970). 
Single photon transition from the 2$^2S$ level to the ground 1$^2S$
is forbidden. Radiative decay of the 2$^2S$ level to the
ground can only occur via two photon emission, in which the electron
jumps to a virtual level, which can lie in anywhere between $n=1$ and 
$n=2$ levels, and then to the ground state, emitting two photons in the
process. As a result, continuum 
emission longward of Ly$\alpha$ will occur. The emission coefficient of the
two-photon process is given by
\begin{equation}
 j_{\nu,2q}=\frac{1}{4\pi}n_{2^2S}hyA(y)~[{\rm erg~cm}^{-3}~{\rm s}^{-1}~{\rm Hz}^{-1}],
\end{equation}
where $y=\nu/\nu_{{\rm Ly}\alpha}$ and $A(y){\rm d}y$ is the probability
per second for emitting a photon in the interval d$y$.
Approximately, the function $A(y)$ is given by the following expression
\begin{equation}
A(y)=202.0\{y(1-y)\{1-[4y(1-y)]^{0.8}\}+0.88[y(1-y)]^{1.53}[4y(1-y)]^{0.8}\} {\rm s}^{-1}
\end{equation}
\citep{kwok}.
The transition probability for the two-photon decay is $A_{2^2S,1^2S}=\int^1_0A(y){\rm d}y=8.23$~s$^{-1}$.
The equilibrium population of the 2$^2S$ state is given by
\begin{equation}
n_{2^2S}=\frac{N_pN_e\alpha^{eff}_{2 ^2S}}{A_{2^2S,1^2S}+N_pq^p_{2^2S,2^2P}+N_eq^e_{2^2S,2^2P}}
\end{equation}
\citep{obsterbrock}, where $N_p$ is the proton density, $\alpha^{eff}_{2^2S}$ is the effective 
recombination coefficient for populating 2$^2S$, and
$q^p_{2^2S, 2^2P}$ and $q^e_{2^2S, 2^2P}$ are the collisional de-excitation
transition rates of 2$^2S$ to 2$^2P$ states by electrons and protons, 
respectively. Note that the collisional transition from 2$^2P$ to 2$^2S$
has been ignored in equation (6). Under Case A conditions, this is a good
approximation. Under Case B conditions, however, the population of the 2$^2P$
state is comparable to that of the metastable 2$^2S$ level since Ly$\alpha$ is trapped between the
2$^2P$ and 1$^2S$ states. Consequently, there is considerable collisional
excitation from 2$^2P$ to 2$^2S$, which enhances the two-photon 
emission. Therefore, the emission coefficient of two-photon process given
by equation (4) can be significantly underestimated. However, this hardly affects our
resultant electron temperatures and densities since an additional continuum
(see below), is added to fit our observed 
spectrum in order to simulate direct and scattered stellar light. For a limited
wavelength region, such an approach is sufficient to account for any excess
emission from the two-photon process. 

Finally, direct or scattered light from the central star
can contaminate the observed continuum level. For our observations, the slit
position was normally chosen to sample the brightest parts of the nebulae,
sometimes passing through the central star. Some spectra were
derived by scanning across the nebular surface. It is quite difficult
to evaluate the level and spectral energy distribution of the contaminating
stellar continuum. As central stars of PNe are
relatively hot ($\ga$20,000\,K), in  our calculation, a Rayleigh-Jeans approximation
for a Black Body ($j_{\lambda,RJ}\varpropto\lambda^{-4}$) is adopted to 
simulate the stellar continuum. This stellar component also partially
compensates for the
underestimated two-photon continuum, as mentioned above. For simplicity, we 
shall call
this term `stellar continuum' hereafter.
Continuum emission from dust grains peaks in the infrared and is negligible in
the optical region. Thus it has been neglected in our calculation.

Apart from continuum emission, recombination line emission from bound-bound transitions
of H$^0$, He$^0$ and He$^+$ is also included. The emission
coefficient of a recombination line is given by
\begin{equation}
 j_{k,bb}=\frac{1}{4\pi}N_eN_i\alpha^{eff}_kh\nu_k~[{\rm erg~cm}^{-3}~{\rm s}^{-1}],
\end{equation}
where $\alpha^{eff}_k$ is the effective recombination coefficient of
the line. For \ion{H}{1} and \ion{He}{2} recombination lines, 
emission coefficients
are determined as in \citet{hummer1987}. Here the considered maximum 
principle quantum number of
is $n=500$. For \ion{He}{1} recombination lines,
emission coefficients are taken from \citet{benjamin} and 
\citet{brocklehurst}. All line profiles 
are assumed Gaussian with a {\it FWHM}
\begin{equation}
\sigma_k=\sqrt{\sigma_{k,{t}}^2+\sigma_{o}^2}~{\rm \AA},
\end{equation}
where $\sigma_{k,{t}}$ is
the  {\it FWHM} caused by thermal broadening, given by 
\begin{equation}
\sigma_{k,t}=2\sqrt{\ln2}\lambda_k(\frac{kT_e}{mc^2})^{1/2} ~{\rm \AA},
\end{equation}
where $m$ is the mass of the atom, and $\sigma_{o}$ is the {\it FWHM} 
caused by other 
broadening effects, such as nebular expansion, turbulence, seeing conditions
and instrument. For a typical PN observed here, 
$\sigma_{o}\ga 8\sigma_{k,{t}}$.
Line emission coefficients are thus converted
into emission coefficients per unit wavelength by
\begin{equation}
j_{\lambda,bb}=\sum_k\frac{1}{\sigma_k\sqrt{2\pi}}{\rm exp}(-\frac{(\lambda-\lambda_k)^2}{2\sigma_k^2})j_{k,bb}~[{\rm erg~cm}^{-3}~{\rm s}^{-1}~{\rm \AA}^{-1}].
\end{equation}
Just longward of the Balmer jump, high-order Balmer lines converge and merge into
a continuum-like feature. The slope of the feature
is sensitive to electron density, and thus provide an excellent
density diagnostic. The region is contaminated by the 
\ion{Ne}{2} $\lambda$3664 line of Multiplet V~1. However, the weak \ion{Ne}{2}
recombination line has a relatively narrow line width compared to the
profile of the continuum-like feature and consequently, cannot cause
any difficulty in our analysis.

%\begin{equation}
%j_{\lambda,tot}=j_{\lambda,ff}({\rm H}^+,{\rm He}^+,{\rm He}^{2+})+j_{\lambda,bf}({\rm H}^+,{\rm He}^+,{\rm He}^{2+})+j_{\lambda,2q}+j_{\lambda,RJ} +\sum_i\frac{1}{\sigma_i\sqrt{2\pi}}exp(-\frac{(\lambda-\lambda_i)^2}{2\sigma^2})j_{\lambda_i,bb}({\rm H}^+,{\rm He}^+,{\rm He}^{2+})~[{\rm erg~cm}^{-3}~{\rm s}^{-1}~\lambda^{-1}], 
%\end{equation}
Summing up contributions from all the processes discussed above, 
the flux at a given wavelength is given by 
\begin{equation}
j_{\lambda,tot}=j_{\lambda,2q}+j_{\lambda,RJ}+\sum_{{\rm X}={\rm H}^0,{\rm He}^0,{\rm He}^+}[j_{\lambda,ff}({\rm X})+j_{\lambda,bf}({\rm X})+j_{\lambda,bb}({\rm X})]~[{\rm erg~cm}^{-3}~{\rm s}^{-1}~{\rm \AA}^{-1}].
\end{equation}

\begin{figure*}
\epsfig{file=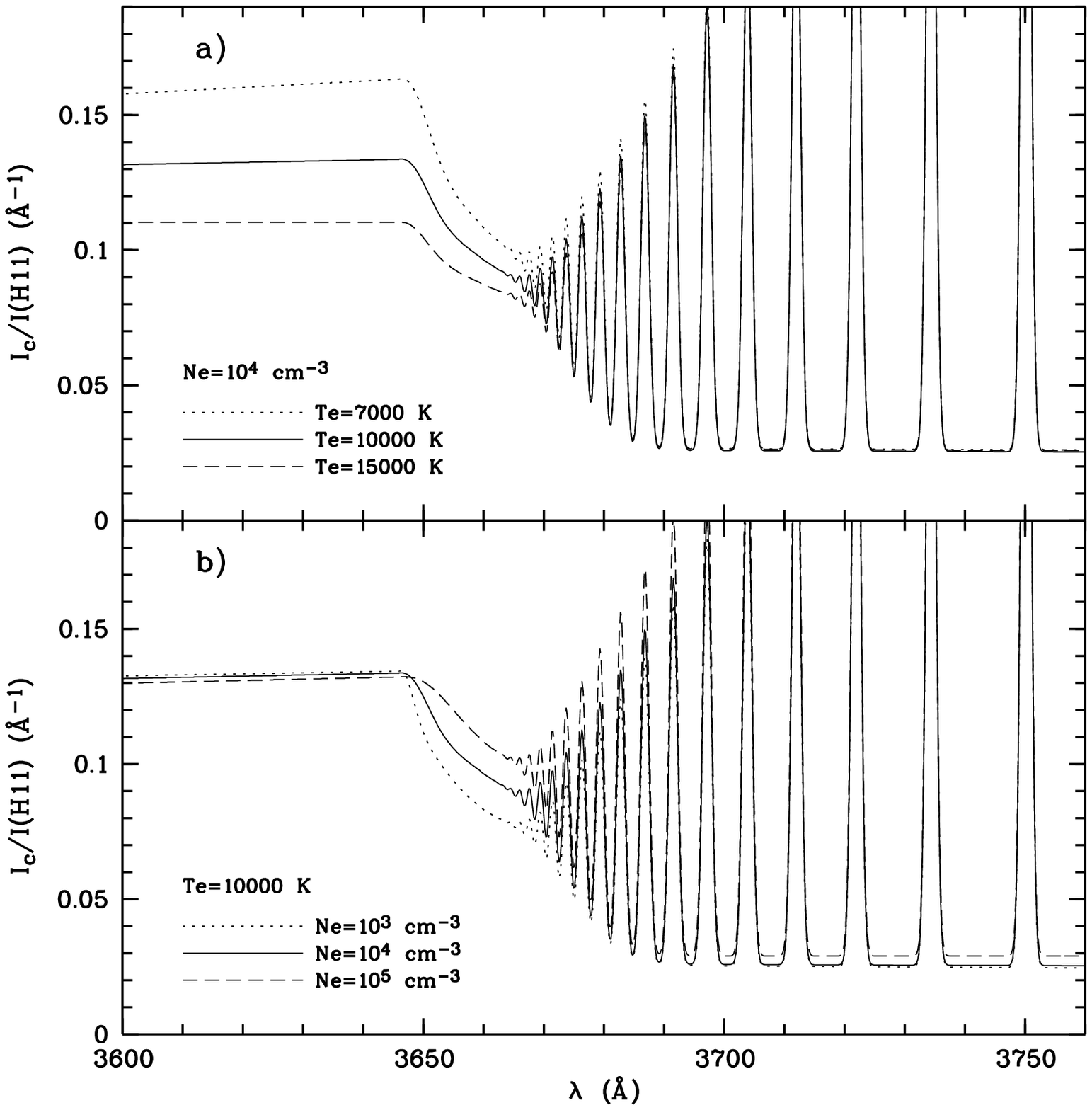,
height=16.5cm, bbllx=42, bblly=72, bburx=539, bbury=578, clip=, angle=0}
\caption{Theoretical hydrogen recombination spectra near the Balmer
discontinuity at (a) same densities but different temperatures and (b) same 
temperatures but different densities.}
\label{trend}
\end{figure*}

Equation (11) shows that, in order to synthesize the nebular spectrum
near the Balmer jump wavelength region, the required input parameters include nebular electron
temperature and density, the magnitude of contaminating stellar continuum,
He$^+$/H$^+$ and He$^{2+}$/H$^+$ ionic abundance ratios and $\sigma_o$.
In general, decreasing (increasing) of the nebular electron temperature will steepen (flatten)
the general shape of the theoretical continuum and increase (decrease) the
Balmer discontinuities; decreasing (increasing) of electron density will
steepen (flatten) the continuous feature redwards of 3646~{\AA} and
decrease (increase) the intensities of high-order Balmer lines relative to a
low-order line, such as H$\beta$, as shown in Fig.~\ref{trend}. Accordingly, matching the dereddened 
observed spectrum to those calculated from equation (11) by varying the
electron temperature and density, we can determine the
electron temperature and density 
simultaneously provided other parameters, such as
He$^+$/H$^+$ and He$^{2+}$/H$^+$ abundance ratios and $\sigma_o$ are known.

\subsection{Results}

Electron temperatures and densities for 48 PNe are derived by fitting the 
synthetic to the observed spectra. For synthesis, 
the required He$^+$/H$^+$ and He$^{2+}$/H$^+$ abundance ratios have
been observationally determined, 
and $\sigma_o$ was derived by fitting the line profiles of strong \ion{H}{1}
recombination lines deconvolved with the thermal broadening, as 
listed in Table~\ref{tbl-1}.
Three free parameters, nebular electron temperature and density and the
fraction of the scattered Rayleigh-Jeans stellar continuum, are optimized simultaneously, using
least Chi square method to match the observed spectra. 
For the optimization, the fitted data points were  
specified explicitly by eye, 
within the wavelength range from 3550--3850\,{\AA}, but
excluding regions of strong (forbidden) lines. 
The results are partially shown in Fig.~\ref{all} together with the residuals of the
fits.

In Table~\ref{tbl-2}, we report the derived
electron temperatures and densities along with their estimated errors.
The Balmer jump and decrement of the \ion{H}{1} recombination spectrum have
previously been used to determine nebular electron temperatures and densities
in four PNe, M~2-24 \citep{zhang2003}, NGC~6153 \citep{liubarlow2000},
M~1-42 and M~2-36 \citep{liu2001}. For the four PNe,
the results derived here 
are in good agreement with the earlier results within the errors.

\subsection{Error analysis}

The errors caused by uncertainties in the input parameters
He$^+$/H$^+$ and He$^{2+}$/H$^+$ abundance ratios are negligible
due to the small measurement errors of strong \ion{He}{1} and
\ion{He}{2} lines and their small contribution to
the continuum emission compared to hydrogen ($\sim10\%$).
Errors in $\sigma_o$ can also be ignored as this quantity is well determined by
fitting strong hydrogen lines.

\begin{figure*}
\epsfig{file=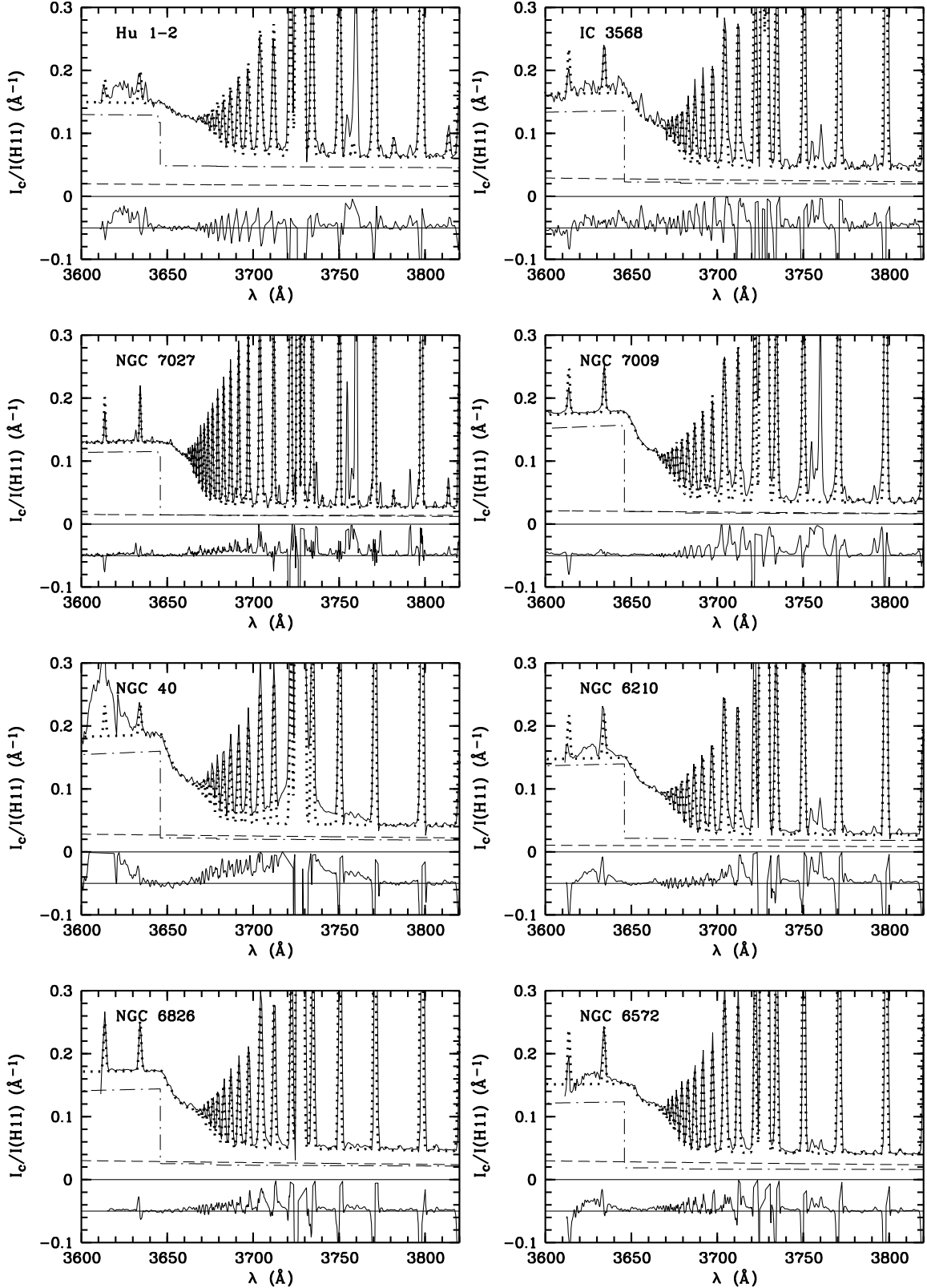,
height=21.5cm, bbllx=28, bblly=43, bburx=548, bbury=771, clip=, angle=0}
\caption{A comparison of observed and theoretical spectra at 3650~\AA. 
The top of each panel shows the 
observed spectrum (solid line) which has been de-redddened using the extinction
constant listed in Table~\ref{tbl-1}, the synthetic spectrum (dotted line), nebular
continuum (dotted-dashed line) and stellar continuum (dashed line). The bottom
shows residuals subtracted by 0.05.
}
\label{all}
\end{figure*}

The complicated velocity 
fields within a nebula can cause different lines to have different deviations 
from their laboratory wavelengths, and thus can affect the quality of line-fitting
made under the simple assumption of a uniform velocity for all ions and lines.
Line-fitting can also be affected by blends of unknown lines and deviations 
of line profiles from Gaussian. 
Inspection of Fig.~\ref{all} shows that the observed
intensities of the \ion{He}{1} 2s\,$^1$S--$5$p\,$^1$P$^{\rm o}$ $\lambda3614$ lines are
generally lower than its predicted values. \citet{liu2001} attributed this
to the destruction of \ion{He}{1} Lyman photons by photoionization of
H$^0$ or by absorption of dust grains. The wavelength region near
the Balmer jump also contains a few strong \ion{O}{3} Bowen fluorescence lines which are
not considered in our simulations, thus giving large residuals at the corresponding wavelengths.
However, these effects do not contribute to errors in the derived electron
temperatures and densities since these wavelengths have been excluded in the
fitting.

The observed spectra plotted in Fig.~\ref{all}, were de-reddened
using extinction constants listed in Table~\ref{tbl-1}.
All spectra were normalized to H~11 $\lambda3770$ in order to minimize errors in derived
temperatures and densities caused by uncertainties
in the reddening correction and flux calibration, and taking the advantage of the small
wavelength difference between the Balmer jump and H~11.

The stellar continuum may
have a spectral energy distribution that deviates from the Rayleigh-Jeans
approximation adopted here. However, this 
continuum is expected to be smooth and vary only slightly 
over such a short wavelength range of fitting. Therefore, we have also
ignored any possible errors caused by the fitting of the stellar continuum.

Errors in the derived temperatures and densities
are dominated by two effects. Noise is an important 
source of error. For most PNe, we found that uncertainties due to this factor
are about 5$\%$ in the derived temperatures and 0.1 dex
in the densities. The other major error source
originates from the extensive wings of the strong 
[\ion{O}{2}] $\lambda\lambda$ 3726, 3729 lines, which can
affect the shape of the continuum near the Balmer jump. 
The effect is larger for low-excitation PNe, such as NGC~40 and NGC~6720.
The extensive wings of the [\ion{O}{2}] lines can cause
the continuum redwards
of the Balmer jump to be overestimated.
As a result, the temperatures can be overestimated.
The errors can be reduced by including data points at longer wavelengths in 
the fitting, 
where the extensive wings of the [\ion{O}{2}] lines have less impact.
However, increasing the wavelength range of the fitting would increase
uncertainties in the reddening correction and in fitting the stellar
continuum. To evaluate errors due to this effect, each spectrum was fitted
twice for two wavelength ranges with one extending further into the red than
the other.
The extensive wings of the [\ion{O}{2}] lines can also decrease the
shape of the continuum-like feature redwards of the Balmer jump, leading to
overestimated densities. For most PNe, this effect is negligible given the
large wavelength difference (about 70~\AA) between the Balmer jump and the 
moderate strength of [\ion{O}{2}] lines. However, the effect can be
significant for those with extremely strong [\ion{O}{2}] lines. On the other hand,
the extensive wings of the [\ion{O}{2}] lines have much less effect on the 
integrated intensities of high-order Balmer lines 
($n\rightarrow2$, $n=11,13,...,23$) which can be determined using a line profile
fitting technique.  
Comparison of densities derived from the continuum-like feature
redwards of the Balmer jump with those deduced from the integrated intensities of 
high-order Balmer lines (e.g. Liu et al. 2001b) allows us to estimate the errors in the derived densities caused by
the extensive wings of the [\ion{O}{2}] lines. The final errors given in
Table~\ref{tbl-2} were based on estimates of contributions from the two
dominating sources of uncertainty discussed above.

\section{Discussion}

\subsection{Comparison of densities}

\begin{figure*}
\epsfig{file=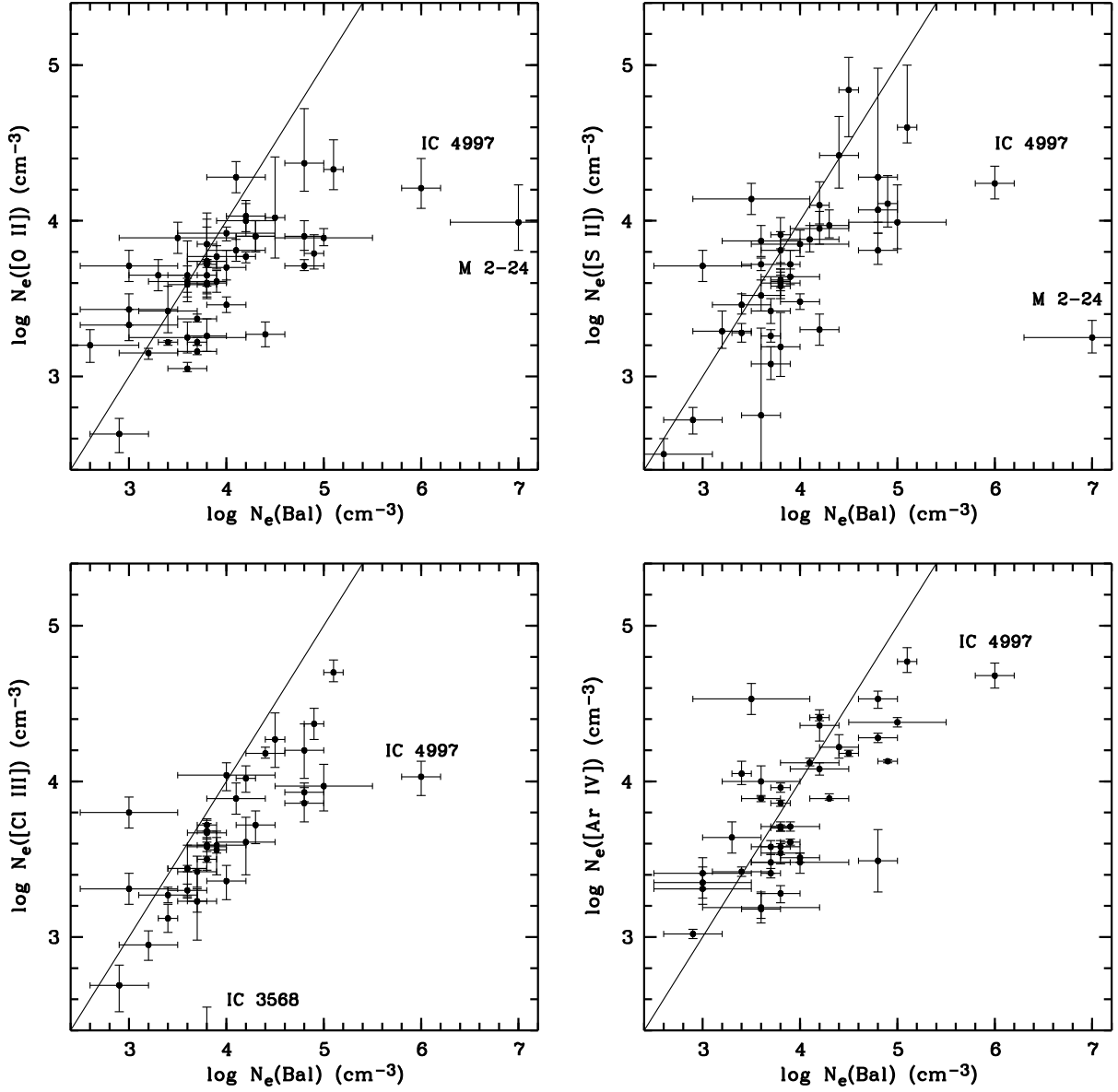,
height=16.5cm, bbllx=28, bblly=252, bburx=548, bbury=772, clip=, angle=0}
\caption{Comparison of the electron densities derived from the intensity
ratios of [\ion{O}{2}] $\lambda3729/\lambda3726$,
[\ion{S}{2}] $\lambda6716/\lambda6731$, [\ion{Cl}{3}] $\lambda5517/\lambda5537$
and [\ion{Ar}{4}] $\lambda4711/\lambda4740$
versus those deduced from the Balmer decrements. The solid lines are $y=x$ plots.}
\label{necorr}
\end{figure*}

\begin{figure*}
\epsfig{file=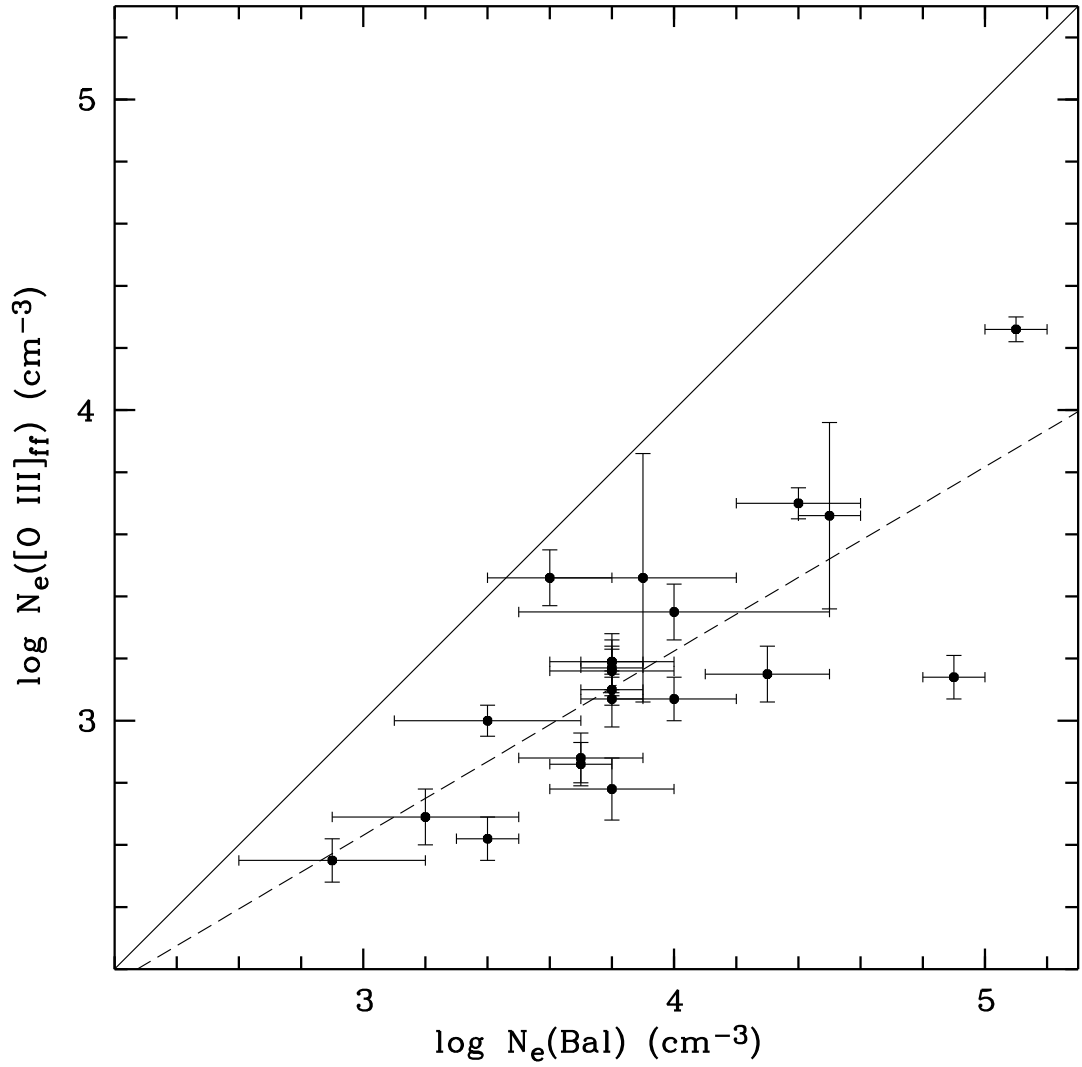,
height=14.5cm, bbllx=72, bblly=156, bburx=524, bbury=605, clip=, angle=0}
\caption{Comparison of the electron densities derived from 
[\ion{O}{3}] $52\mu{\rm m}/88\mu{\rm m}$ ratios
and the Balmer decrements. The solid line is a $y=x$ plot and the dashed line
is a linear fit.}
\label{necorrir}
\end{figure*}

Electron densities deduced from the Balmer decrements, and from the
[\ion{O}{2}] $\lambda3729/\lambda3726$,
[\ion{S}{2}] $\lambda6716/\lambda6731$, [\ion{Cl}{3}] $\lambda5517/\lambda5537$
and [\ion{Ar}{4}] $\lambda4711/\lambda4740$ doublet ratios are compared
in Fig.~\ref{necorr}. Fig.~\ref{necorr} shows that for most PNe, the 
hydrogen recombination spectrum yields
higher densities compared to forbidden
lines diagnostics. There is a trend such that the discrepancy becomes
larger at higher densities. Two extreme cases are IC~4997 and M~2-24.
For IC~3568, [\ion{Cl}{3}] $\lambda5517/\lambda5537$ yields an abnormally
low density. The cause of it is unclear but is probably due to measurement
error. In Fig.~\ref{necorrir} we compare
electron densities derived from the [\ion{O}{3}] far-infrared fine
structure line ratio
and from the hydrogen recombination spectrum. Evidently, densities derived
from the [\ion{O}{3}] $52\mu{\rm m}/88\mu{\rm m}$ ratio are significantly
lower than those derived from the Balmer decrements.
In fact, it has been found that the far-IR lines yield systematic lower
densities, lower than even those deduced from optical forbidden line
density-diagnostics \citep{liubarlow2001}. A linear fit
shows that the density discrepancy increases with the electron
density.

O$^0$, S$^0$, Cl$^{+}$, O$^{+}$ and Ar$^{2+}$ have ionization potentials of
13.6, 10.4, 23.8, 35.1 and 40.7~eV, respectively, and therefore embrace a wide 
range of regions of differing states of ionization. 
 On the other hand, the hydrogen recombination spectrum arises from
the entire ionized nebula. Thus the discrepancies between densities derived
from \ion{H}{1} recombination spectrum and those deduced from forbidden line
ratios are not likely to be caused by ionization stratifications. 
We notice that the
[\ion{O}{2}], [\ion{S}{2}], [\ion{Cl}{3}], [\ion{Ar}{4}] and [\ion{O}{3}]
density-diagnostic
lines have critical densities of $N_{\rm crit}\la10^5$~cm$^{-3}$.
Therefore, if the nebulae contain regions of higher densities,
these forbidden lines will be suppressed in these regions by collisional 
de-excitation. In contrast, the hydrogen recombination spectrum is practically 
unaffected by collisional de-excitation.
As a result, in such cases the hydrogen recombination spectrum
will yield higher average densities compared to the
[\ion{O}{2}], [\ion{S}{2}], [\ion{Cl}{3}], [\ion{Ar}{4}] and [\ion{O}{3}]
lines ratios. The same argument has been developed by \citep{liubarlow2001} 
to explain the systematically lower densities derived from the [\ion{O}{3}]
fine-structure line ratio compare to those deduced from optical forbidden line
ratios. The [\ion{Ar}{4}] lines have higher critical
densities than the [\ion{O}{2}], [\ion{S}{2}], [\ion{Cl}{3}] lines. Thus,
the effects are weaker for [\ion{Ar}{4}] doublet ratio than for the other
cases. Given that the [\ion{O}{3}] far-infrared fine-structure lines have the lowest
critical densities amongst those density-diagnostic forbidden lines
($N_{\rm cri}$([\ion{O}{3}]$_{\rm ff})\la4\times10^3$~cm$^{\rm -3}$),
the effects are largest for the [\ion{O}{3}] 
$52\mu{\rm m}/88\mu{\rm m}$ ratio. This can be seen clearly in 
Fig.~\ref{necorr} and Fig.~\ref{necorrir}.
A nebular
model including these high-density condensations and their effects on temperature
determinations will be discussed in Section 4.3.

\subsection{Comparison of temperatures}
\begin{figure*}
\epsfig{file=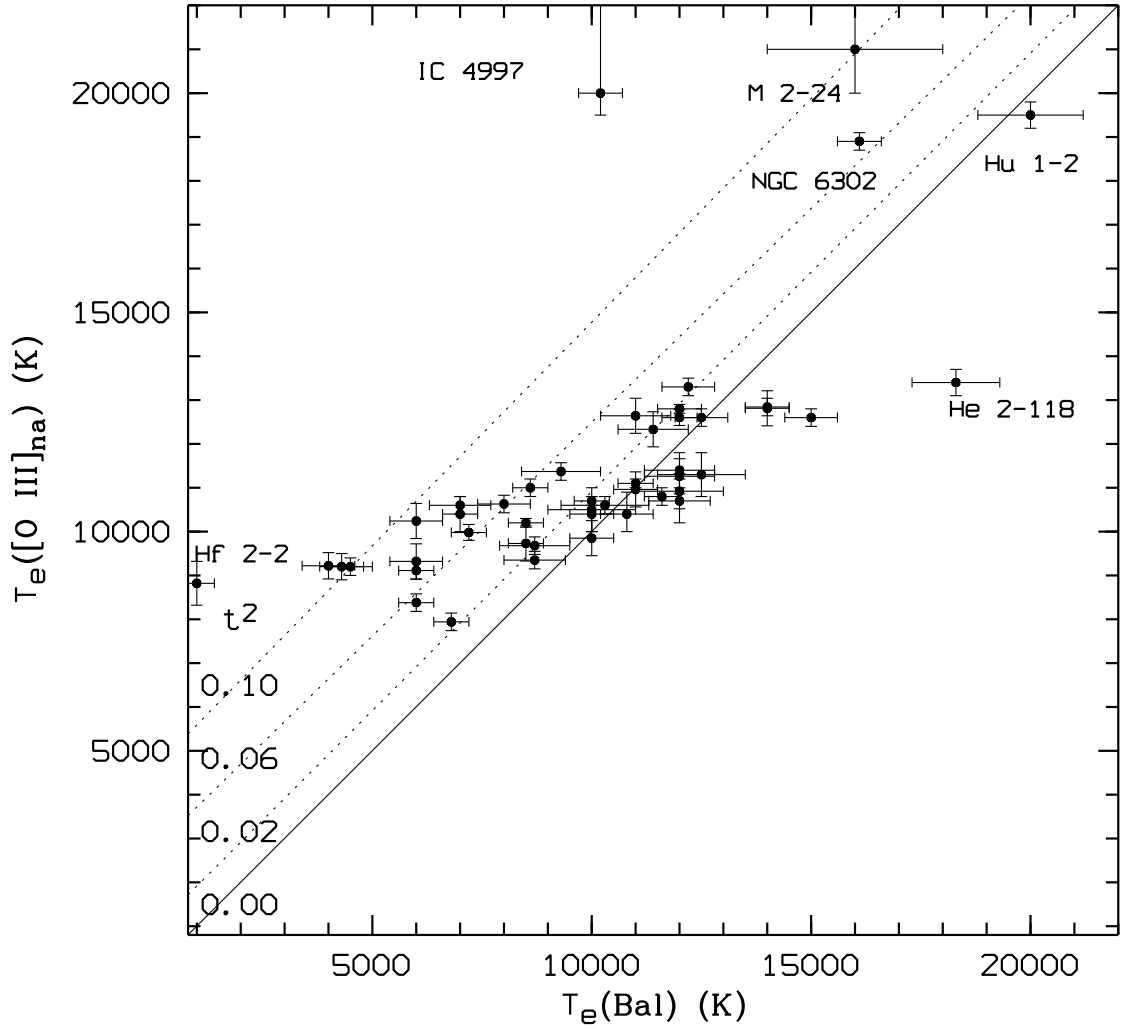,
height=14cm, bbllx=39, bblly=156, bburx=522, bbury=605, clip=, angle=0}
\caption{Comparison of the electron temperatures derived from the 
nebular-to-auroral line ratio of [\ion{O}{3}] and those deduced from
the Balmer decrements. The lines show the variation of 
$T_{\rm e}$([\ion{O}{3}]$_{\rm na}$) as a function of $T_{\rm e}$(Bal)
for the mean square temperature fluctuation $t^2=0.00, 0.02, 0.06$ and 0.10.}
\label{tecorr}
\end{figure*}

\begin{figure*}
\epsfig{file= 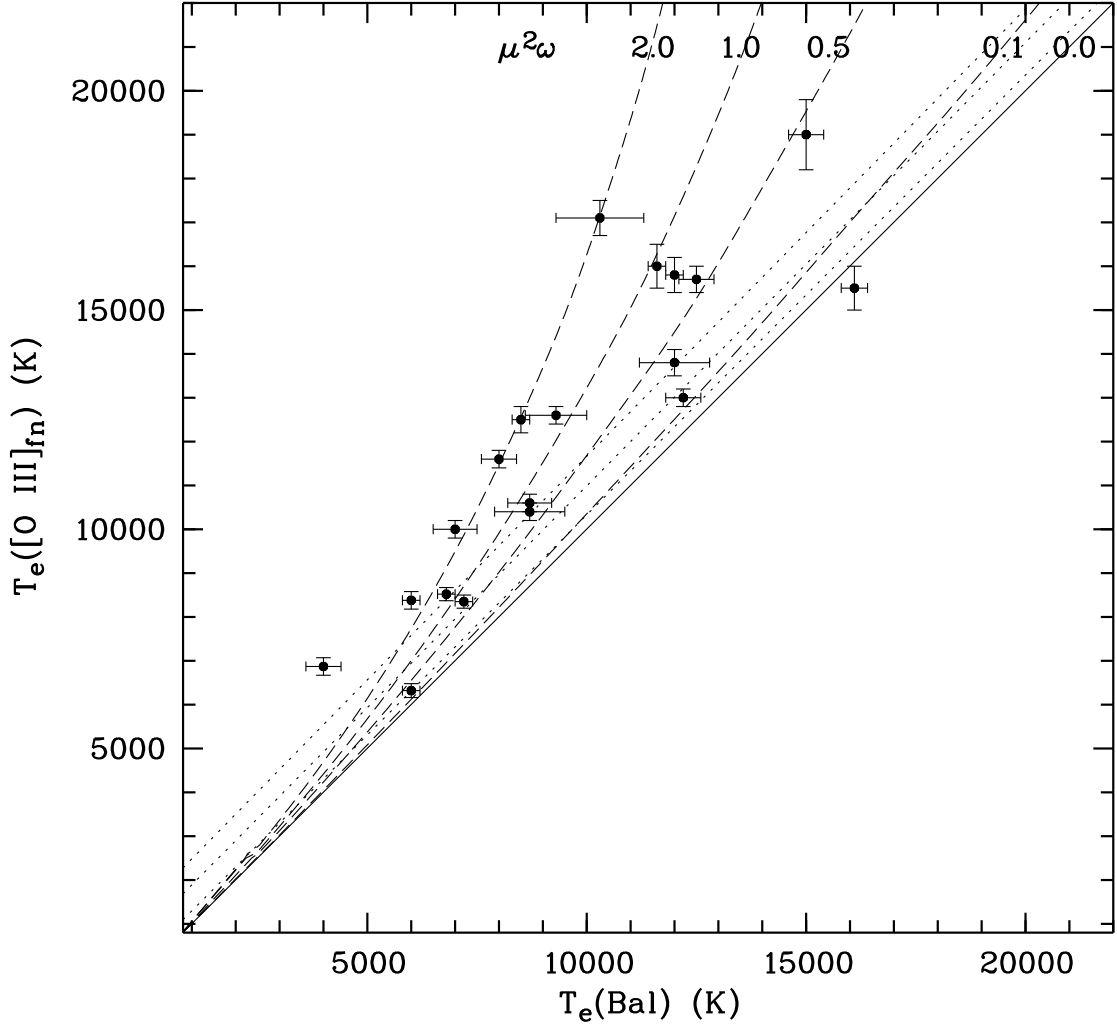,
height=14cm, bbllx=39, bblly=156, bburx=522, bbury=605, clip=, angle=0}
\caption{Comparison of the electron temperatures derived from the
[\ion{O}{3}] $(52\mu{\rm m}+88\mu{\rm m})/\lambda4959$ ratios and those deduced from
the Balmer decrements. The dotted lines show variation of
$T_{\rm e}$([\ion{O}{3}]$_{\rm fn}$) as a function of $T_{\rm e}$(Bal)
for the mean square temperature fluctuation $t^2=0.02, 0.06$ and 0.10.
The dashed lines show variation of
$T_{\rm e}$([\ion{O}{3}]$_{\rm fn}$) as a function of $T_{\rm e}$(Bal)
for $\mu^2\omega=0.1, 0.5, 1.0$ and 2.0 (see Section~4.2). The solid line
shows that for homogeneous temperature and density.}
\label{tecorrir}
\end{figure*}

Fig.~\ref{tecorr} and Fig.~\ref{tecorrir} show that electron
temperatures derived from the [\ion{O}{3}] $\lambda4959/\lambda4363$ and
$(52\mu{\rm m}+88\mu{\rm m})/\lambda4959$ ratios 
[hereafter, $T_{\rm e}$([\ion{O}{3}]$_{\rm na})$ and $T_{\rm e}$([\ion{O}{3}]$_{\rm fn})$] are
systematically higher than those deduced from the hydrogen recombination spectrum [hereafter $T_{\rm e}$(Bal)],
although there are a few exceptions where $T_{\rm e}$(Bal) is higher 
than $T_{\rm e}$([\ion{O}{3}]$_{\rm na}$).
The most extreme case is He~2-118, for which $T_{\rm e}$(Bal) is 5000~K higher
than $T_{\rm e}$([\ion{O}{3}]$_{\rm na}$).
In our sample, the four PNe with the largest temperatures discrepancies,
Hf~2-2, M~1-42, M~2-36 and NGC~6153, were previously studied in detail by 
\citet{liu209}, \citet{liu2001} and \citet{liubarlow2000}.

The discrepancy between $T_{\rm e}$([\ion{O}{3}]$_{\rm na}$) and 
$T_{\rm e}$(Bal) was first discovered by \citet{peim71} and was 
attributed to temperature fluctuations within the nebula \citep{peim67}. 
He characterized the temperature structure
of a nebula by the average temperature $T_0$ and a mean square temperature
fluctuation $t^2$ as follows:
\begin{equation}
T_0(N_{\rm e},N_i)=\frac{\int T_{\rm e}N_{\rm e}N_idV}{\int N_{\rm e}N_idV}
\end{equation}
and
\begin{equation}
t^2=\frac{\int (T_{\rm e}-T_0)^2N_{\rm e}N_idV}{T_0^2\int N_{\rm e}N_idV},
\end{equation}
where $N_i$ is the ion density for an observed emission line. Assuming 
that all the oxygen is twice ionized and that there are no density and
composition variations in the nebula, $T_0$ and $t^2$ can
be derived from two of the three expressions,
\begin{equation}
T_{\rm e}([{\rm O}~{\rm III}]_{\rm na})=T_0[1+\frac{1}{2}(\frac{9.13\times10^4}{T_0}-3)t^2],
\end{equation}
\begin{equation}
T_{\rm e}([{\rm O}~{\rm III}]_{\rm fn})=T_0[1+\frac{1}{2}(\frac{2.92\times10^4}{T_0}-3)t^2],
\end{equation}
and 
\begin{equation}
T_{\rm e}({\rm Bal})=T_0(1-1.67t^2),
\end{equation}
\citep{peim67,dinerstein}.

In Fig.~\ref{tecorr} and Fig.~\ref{tecorrir}, lines showing $T_{\rm e}$([\ion{O}{3}]$_{\rm na}$) as a  
function of $T_{\rm e}({\rm Bal})$ for the case of $t^2=0.00$, 0.02, 0.06 and 1.00
are over plotted. Fig.~\ref{tecorr} shows that most nebulae have $t^2\la0.06$. 
Excluding a few extreme nebulae having $t^2>0.1$, we obtain a mean value of $<t^2>=0.031$.
The value is in excellent agreement with previous result $<t^2>=0.03$ derived by \citet{liu93} although
their analysis contained larger measurement errors. However, even excluding the
extreme nebulae, a similar comparison between 
$T_{\rm e}$([\ion{O}{3}]$_{\rm fn}$) and $T_{\rm e}({\rm Bal})$ yields an
average $<t^2>=0.162$. 
As we shall show in the following subsection, it is likely that $T_{\rm e}$([\ion{O}{3}]$_{\rm fn}$)
have been overestimated because of density variations in the nebula, resulting
in higher $<t^2>$ when comparing $T_{\rm e}$([\ion{O}{3}]$_{\rm fn}$) and $T_{\rm e}({\rm Bal})$.

Another interesting point revealed by Fig.~\ref{tecorr} is that
$T_{\rm e}({\rm Bal})$ covers a wider range of values than 
$T_{\rm e}$([\ion{O}{3}]$_{\rm na}$). Except a few extreme cases, 
$T_{\rm e}$([\ion{O}{3}]$_{\rm na}$) falls between 8000 and 14000~K,
whereas $T_{\rm e}({\rm Bal})$ varies from 900 to 19000~K.
\citet{peim67} has argued that temperature fluctuations of gaseous nebulae
can cause lower $T_{\rm e}({\rm Bal})$ compared to
$T_{\rm e}$([\ion{O}{3}]$_{\rm na}$). However, even $<t^2>=0.031$ is beyond
any photoionization model prediction. Another explanation of higher 
$T_{\rm e}({\rm Bal})$ compared to $T_{\rm e}$([\ion{O}{3}]$_{\rm na}$)
is that there are some condensations with $N_{\rm e}>10^6$~cm$^{-3}$ in PNe,
which then suppress the $T_{\rm e}$([\ion{O}{3}]) $\lambda\lambda$4959,5007 
relative to the $\lambda$4363, leading to an artificially higher 
$T_{\rm e}$([\ion{O}{3}]$_{\rm na}$) \citep{viegas}.
But no such high density condensations have been found in PNe having extremely
large temperature discrepancies \citep{liubarlow2000,liu2001}.
Recently, \citet{liubarlow2000} presented an empirical nebular model
containing two components, each with its own temperature, density and
chemical composition to account for the temperature discrepancy. This seems
to be the most plausible explanation at present.

For PNe having $T_{\rm e}({\rm Bal})>T_{\rm e}$([\ion{O}{3}]$_{\rm na}$),
as in He~2-118, the electron temperature might  conceivably increase in 
low-ionization regions owing to heating by shock waves in the outer regions.
As a result, the average temperature of the entire nebula
[$T_{\rm e}$(Bal)] appears higher 
than that of high-ionization regions [$T_{\rm e}$([\ion{O}{3}]$_{\rm na}$)]. 

\subsection{A simple model with condensations}

In \S~4.1, we found that electron densities derived from the \ion{H}{1} 
recombination spectrum are systematically higher than those derived from
optical forbidden lines ratios, and confirm the earlier result of
\citet{liubarlow2001} that densities derived from the [\ion{O}{3}]
far-IR fine-structure lines are the lowest of all. These results show
that nebular electron densities derived from various diagnostics are correlated
with the critical density of the diagnostic lines, and suggest that dense clumps are generally present
in PNe. 

In such cases, the [\ion{O}{3}] 
$(52\micron+88\micron)/\lambda4959$ ratio is no longer a good temperature indicator
given the relatively low critical densities of the [\ion{O}{3}]
$52\micron$- and 88-$\micron$ 
lines (3500 and 1500~cm$^{-3}$, respectively). When nebulae contain
condensations with densities in the range between 10$^4$ and
10$^6$~cm$^{-3}$, the [\ion{O}{3}] $52\micron$- and 88-$\micron$ 
far-infrared lines will be heavily suppressed by collisional de-excitation, 
leading to overestimates of $T_{\rm e}$([\ion{O}{3}]$_{\rm fn}$). On the other 
hand, condensations of densities in this range will not affect
$T_{\rm e}$([\ion{O}{3}]$_{\rm na}$) given that the
[\ion{O}{3}] $\lambda4959$ 
and $\lambda4363$ lines have higher critical densities of
$7\times10^5$~cm$^{-3}$ and $3\times10^7$~cm$^{-3}$, respectively (cf. Fig~~\ref{tecorrir}).

\begin{figure*}
\epsfig{file= 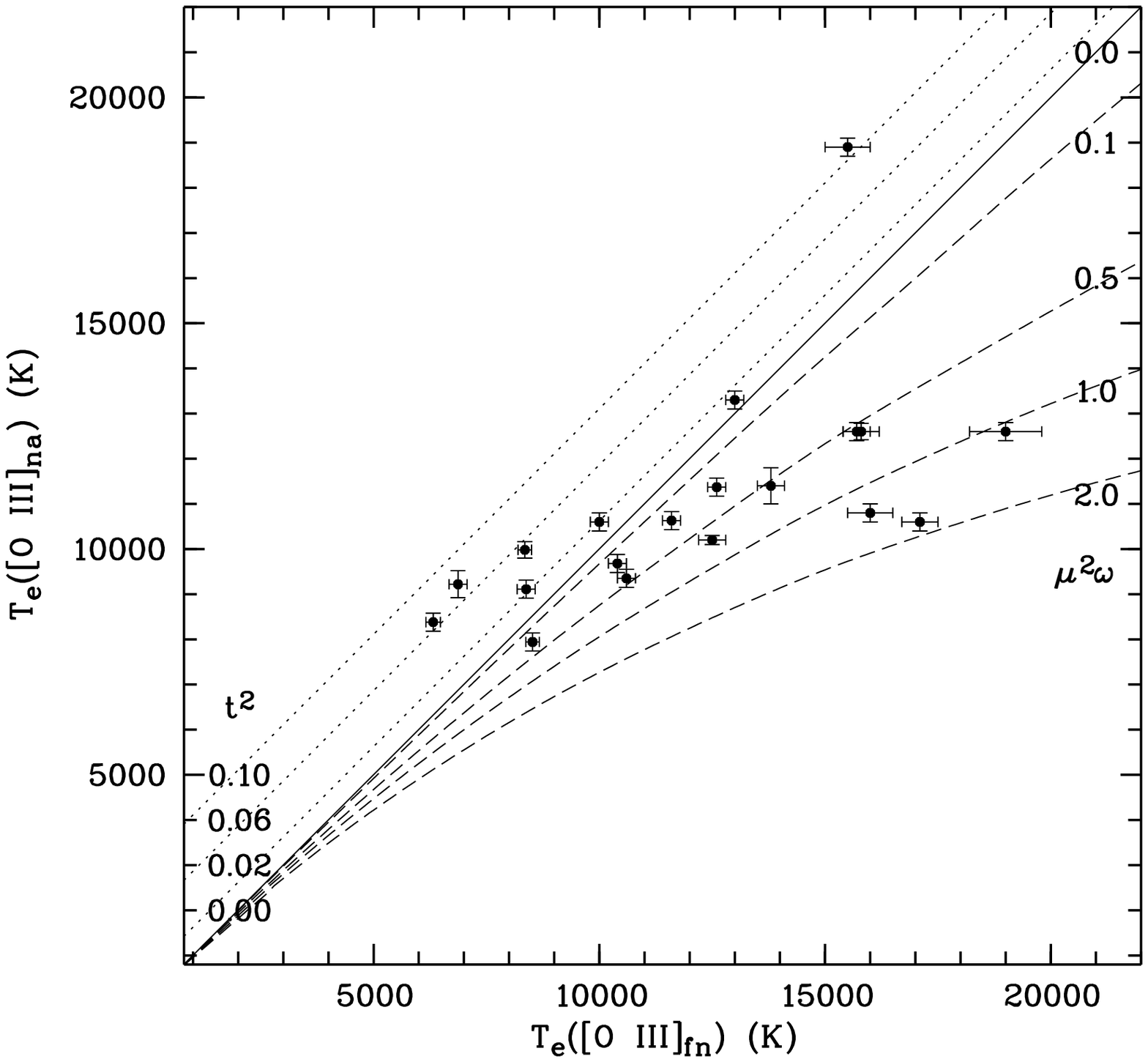,
height=14cm, bbllx=39, bblly=156, bburx=527, bbury=605, clip=, angle=0}
\caption{Comparison of the electron temperatures derived from the
[\ion{O}{3}] $\lambda4959/\lambda4363$ ratios and those deduced
from
the [\ion{O}{3}] $(52\mu{\rm m}+88\mu{\rm m})/\lambda4959$ ratios. The dotted lines show the variation of
$T_{\rm e}$([\ion{O}{3}]$_{\rm na}$) as a function of $T_{\rm e}$([\ion{O}{3}]$_{\rm fn}$)
for the mean square temperature fluctuation $t^2=0.02, 0.06$ and 0.10.
The dashed lines show the variation of
$T_{\rm e}$([\ion{O}{3}]$_{\rm na}$) as a function of $T_{\rm e}$([\ion{O}{3}]$_{\rm fn}$)
for $\mu^2\omega=0.1, 0.5, 1.0$ and 2.0. The solid line
shows that for homogeneous temperature and density.}
\label{oiiite}
\end{figure*}

Another piece of evidence pointing to the presence of dense clumps comes
from Fig.~\ref{oiiite} where we compare
$T_{\rm e}$([\ion{O}{3}]$_{\rm na}$) and $T_{\rm e}$([\ion{O}{3}]$_{\rm fn}$).
If nebulae have temperature fluctuations but are homogeneous in density,
then $T_{\rm e}$([\ion{O}{3}]$_{\rm fn}$) is expected to be always lower than
$T_{\rm e}$([\ion{O}{3}]$_{\rm na}$) \citep{dinerstein}. However,  Fig.~\ref{oiiite}
shows about two-third of the objects have higher
$T_{\rm e}$([\ion{O}{3}]$_{\rm fn}$) compared to $T_{\rm e}$([\ion{O}{3}]$_{\rm na}$),
suggesting that density
variations play an important role in $T_{\rm e}$([\ion{O}{3}]$_{\rm fn}$)
determinations.

\citet{viegas} showed that electron temperatures derived from the
[\ion{O}{3}] $\lambda4363/(\lambda\lambda4959+5007)$ ratio may be
overestimated if nebulae contain ionized condensations of electron densities 
higher than 
10$^6$~cm$^{-3}$. However, except for a few extreme PNe, such as IC~4997
\citep{hyung1994},
Mz~3 \citep{zhang02} and M~2-24 \citep{zhang2003}
there is no evidence that nebulae contain a
substantial amount of ionized gas in condensations with densities in excess
of 10$^6$~cm$^{-3}$, as the presence of such condensations should have been
revealed by the \ion{H}{1} recombination spectrum (Liu et al. 2000, 2001a, 2001b; Fig.~\ref{necorr}).
Therefore, we consider $N_{\rm e}\sim10^5$~cm$^{-3}$ 
as a safe upper limit for the density of these condensations. 

In order to estimate the filling factor of the condensations, we use an
analytical method originally developed by \citet{viegas} for explaining the discrepancy
of $T_{\rm e}$([\ion{O}{3}]$_{\rm na}$) and $T_{\rm e}({\rm Bal})$. 
Assuming that some high-density regions ($N_{\rm e}>10^6$~cm$^{-3}$) 
are present, they deduced nebular-to-auroral [\ion{O}{3}] line ratios (see
their Eq.~2.3). Following
the same deduction of their Eq.~2.3 but assuming that the electron density of these
 condensations is about
$10^5$~cm$^{-3}$ instead of $10^6$~cm$^{-3}$, we obtain
the infrared-to-auroral line ratio
\begin{equation}
\frac{I_{52\micron+88\micron}}{I_{4959}}=\frac{\varepsilon_{\rm fL}}{\varepsilon_{\rm nL}}\frac{1+\mu^2\omega(\varepsilon_{\rm fH}/\varepsilon_{\rm fL})}{1+\mu^2\omega(\varepsilon_{\rm nH}/\varepsilon_{\rm nL})},
\end{equation} 
where $\varepsilon$ is line emission coefficient, 
$4\pi j=n_{\rm e}n_i\varepsilon$, f and n refer to IR fine-structure line 
and nebular line, L and H refer to the low- and
high-density regions, $\mu=(N_{\rm e})_{\rm H}/(N_{\rm e})_{\rm L}$ is the
density contrast and $\omega=V_{\rm H}/V_{\rm L}$ is the filling factor.
In the following calculation, we assume that: 1) density variations are the 
only cause of the
lower $T_{\rm e}$([\ion{O}{3}]$_{\rm fn}$) compared to
$T_{\rm e}$([\ion{O}{3}]$_{\rm na}$) and $T_{\rm e}(\rm Bal)$; 2) the nebular
gas has a homogeneous electron temperature, i.e., 
$T_{\rm e}$([\ion{O}{3}]$_{\rm na}$) or $T_{\rm e}(\rm Bal)$; 3) the 
condensations have
densities lower than the critical density of the [\ion{O}{3}]
auroral line but higher than the critical densities of the [\ion{O}{3}]
infrared lines, whereas low-density regions have a density that is lower than 
the critical densities of the [\ion{O}{3}] auroral and infrared lines.

Setting $F(T_{[{\rm O}~{\rm III}]_{\rm fn}})=I_{52\micron+88\micron}/I_{4959}$,
we have $\varepsilon_{\rm fL}/\varepsilon_{\rm nL}=F(T_{\rm Bal})$.
For a simple assumption 
$(N_{\rm e})_{\rm H}=10^5$~cm$^{-3}$ and 
$(N_{\rm e})_{\rm L}=10^3$~cm$^{-3}$, the ratio 
$\varepsilon_{\rm fH}/\varepsilon_{\rm fL}$ is about 0.04, whereas 
$\varepsilon_{\rm n}$ is practically independent of the density when
the densities are lower than the critical density of the nebular line and thus
$\varepsilon_{\rm nH}/\varepsilon_{\rm nL}\approx1$.  Therefore, $\mu^2\omega$ 
is given by
\begin{equation}
\mu^2\omega=\frac{F(T_{[{\rm O}~{\rm III}]_{\rm fn}})/F(T_{\rm Bal})-1}{0.04-F(T_{[{\rm O}~{\rm III}]_{\rm fn}})/F(T_{\rm Bal})}.
\end{equation}

In Fig.~\ref{tecorrir}, the lines showing the variation of $T_{\rm e}$([\ion{O}{3}]$_{\rm fn}$) as a
function of $T_{\rm e}$(Bal) for $\mu^2\omega=0.0$, 0.1, 0.5, 1.0 and 2.0
are plotted. We obtain an average value of
$<\mu^2\omega>=1.0$. However, it should be borne in mind that the $\mu^2\omega$
derived by eq.~(18) is a upper limit since temperature variations are actually
present and contribute partly to the measured temperature discrepancies. 

$\mu^2\omega$ can also be estimated by comparing 
$T_{\rm e}$([\ion{O}{3}]$_{\rm na}$) with $T_{\rm e}$([\ion{O}{3}]$_{\rm fn}$),
as shown in Fig.~\ref{oiiite}. 
Assuming that $T_{\rm e}$([\ion{O}{3}]$_{\rm na}$) is constant across the
entire nebula,
$\mu^2\omega$ is likewise given by
\begin{equation}
\mu^2\omega=\frac{F(T_{[{\rm O}~{\rm III}]_{\rm fn}})/F(T_{[{\rm O}~{\rm III}]_{\rm na}})-1}{0.04-F(T_{[{\rm O}~{\rm III}]_{\rm fn}})/F(T_{[{\rm O}~{\rm III}]_{\rm na}})}.
\end{equation} 
In Fig.~\ref{oiiite}, the lines showing the variation of $T_{\rm e}$([\ion{O}{3}]$_{\rm fn}$) as a
function of $T_{\rm e}$(Bal) for $\mu^2\omega=0.0$, 0.1, 0.5, 1.0 and 2.0
are also plotted. An average value of $<\mu^2\omega>=0.4$ is obtained.
It is noteworthy that the presence of temperature variations will increase
$T_{\rm e}$([\ion{O}{3}]${\rm na}$) with respect to 
$T_{\rm e}$([\ion{O}{3}]$_{\rm fn}$), in contrast to density variations. 
Therefore, $\mu^2\omega$ derived from eq.~(19) is a lower limit.

The upper- and lower-limit of $\mu^2\omega$ are listed in Table.~\ref{tbl-3b}.
For a few PNe having a lower limit $(\mu^2\omega)_{\rm L}=0.0$, 
the effect of temperature variations on the temperature
determination is larger than that of density variations, causing higher
$T_{\rm e}$([\ion{O}{3}]$_{\rm na})$ compared to $T_{\rm e}$([\ion{O}{3}]$_{\rm fn}$). 
For all the PNe, the mean value of $\mu^2\omega$ is in the range from 0.4 to 1.0. Thus, we suggest that $\mu^2\omega=0.7$
may be a representative value that can be used for a typical nebula. Adopting
the density contrast $\mu\sim100$, a representative value of the condensations
filling factor $\omega$ is about $7\times10^{-5}$. The value is 
similar to that found by \citet{viegas}.
Adopting $\mu^2\omega=0.7$ for all the PNe, we obtain 
new $T_{\rm e}$([\ion{O}{3}]$_{\rm fn}$), for which the contribution of the
clumps has been deduced using equation~(17). Comparing the new 
$T_{\rm e}$([\ion{O}{3}]$_{\rm fn}$) and $T_{\rm e}$(Bal), we 
re-evaluate the average value $<t^2>=0.032$, which is in good agreement with
the value yielded by the comparison of $T_{\rm e}$([\ion{O}{3}]$_{\rm na}$) 
and $T_{\rm e}$(Bal) (see Section~4.2). 

\subsection{The difference between temperatures derived from the
[\ion{O}{3}] forbidden lines and from the Balmer jump versus electron density}

\begin{figure*}
\epsfig{file= 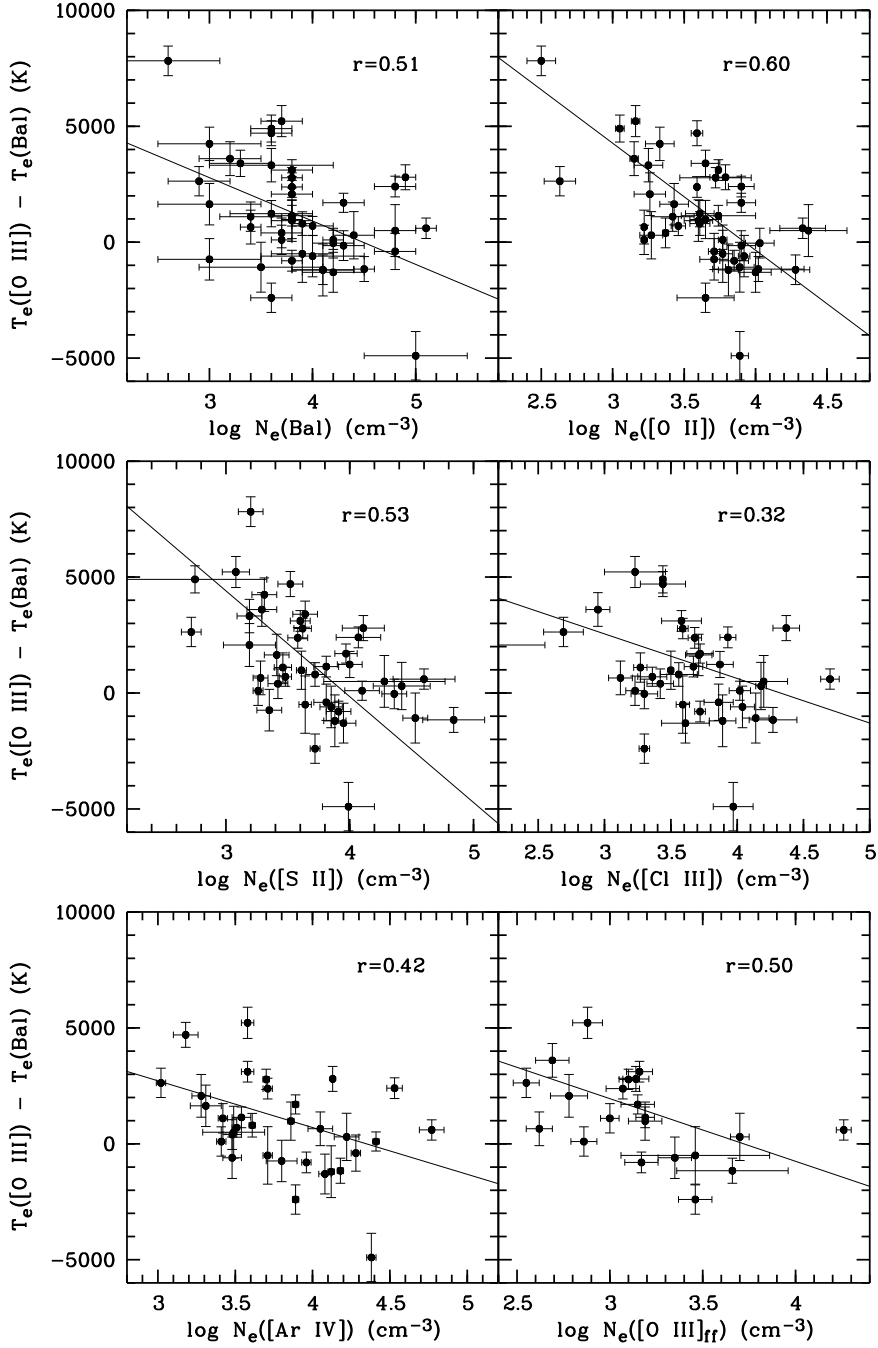,
height=18cm, bbllx=53, bblly=51, bburx=522, bbury=774, clip=, angle=0}
\caption{ $T_{\rm e}$([\ion{O}{3}])$ - T_{\rm e}$(Bal) versus 
$N_{\rm e}$(Bal), $N_{\rm e}$([\ion{O}{2}]), $N_{\rm e}$([\ion{S}{2}]),
$N_{\rm e}$([\ion{Cl}{3}]), $N_{\rm e}$([\ion{Ar}{4}]) and 
$N_{\rm e}$([\ion{O}{3}$_{\rm ff}$]). The solid lines are
least-squares fits. The linear correlation coefficient ``r'' is given
in every panel. For the fits the observed points are weighted according
to the error bar for each. For all the
cases, a trend of increasing discrepancy of temperature derived from
[\ion{O}{3}] forbidden lines and the Balmer jump with decreasing electron
density is seen.}
\label{crene}
\end{figure*}

In Fig.~\ref{crene}, the differences between temperatures derived from
the [\ion{O}{3}] forbidden lines and from the Balmer jump are plotted against 
electron densities derived from the hydrogen recombination spectrum, 
the [\ion{O}{2}] $\lambda3729/\lambda3726$,
[\ion{S}{2}] $\lambda6716/\lambda6731$, [\ion{Cl}{3}] $\lambda5517/\lambda5537$
[\ion{Ar}{4}] $\lambda4711/\lambda4740$ and the [\ion{O}{3}] $52\mu$m/$88\mu$m ratios, 
respectively.  In all cases, the differences between the two temperature
are anti-correlated with nebular densities derived from
various density indicators. A least-squares fit yields the following
relation:
\begin{equation}
\Delta T=(1.30\pm0.12)\times10^4-(2.93\pm0.27)\times10^3\log N_{\rm e},
\end{equation}
and a linear correlation coefficient of 0.51,
where $\Delta T=T_{\rm e}$([\ion{O}{3}])$-T_{\rm e}({\rm Bal})$ and
$N_{\rm e}$ is that derived from hydrogen recombination spectrum, which 
represents the average density of the entire nebula.
\citet{garnett} found that the discrepancies of O$^{2+}$/H$^{+}$ abundance
ratios derived from ORLs and from CELs are greater for larger, lower-surface brightness
PNe than for compact, dense ones. On the other hand, \citet{liu2001} found
that the abundance discrepancies are greater for those with larger
temperature discrepancies. Our results are therefore
consistent with these earlier findings. 
\citet{liu2001} found that,
\begin{equation}
\Delta({\rm O}^{2+}/{\rm H}^+)=(0.209\pm0.085)+(2.01\pm0.33)\times10^{-4}\Delta T,
\end{equation}
where $\Delta({\rm O}^{2+}/{\rm H}^+)=\log({\rm O}^{2+}/{\rm H}^+)_{\rm ORL}-\log({\rm O}^{2+}/{\rm H}^+)_{\rm CEL}$.
From the two relations given in equations (20) and (21), we find that the ratio of abundances derived from ORLs and from CELs has a
power-law dependence on density,
\begin{equation}
\frac{{\rm O}^{2+}_{\rm ORL}}{{\rm O}^{2+}_{\rm CEL}}\propto N_{\rm e}^{-\alpha},
\end{equation}
where $\alpha=0.59\pm0.15$.

In general, electron densities of PNe are expected to decline monotonically
as nebulae expand.
\citet{sk1987} found an approximate relation between nebular mean electron 
density and nebular size, $N_{\rm e}\propto R^{-3}$, where R is the nebular
radius. 
Consequently, it is suggested
that the temperature discrepancy is related to the evolutionary state of PNe;
more expended PNe show larger temperature discrepancies. On the other hand,
the actual evolution in a given nebula may depend on many factors. Furthermore,
the electron density measured for a given nebula may depend on the 
diagnostic used. For example,
electron densities of low-ionization regions, as probed by 
the [\ion{O}{2}] $\lambda3729/\lambda3726$ and 
[\ion{S}{2}] $\lambda6716/\lambda6731$ line ratios, can be enhanced by
interactions of fast stellar winds with earlier slow winds. Therefore,
there are slight difference in deriving the relation of $\Delta T$
and $N_{\rm e}$ from different density diagnostics, as shown in 
Fig.~\ref{crene}. We also use electron density derived from 
these forbidden line ratios to estimate the $\alpha$-value in equation (22).
The results show that the value is within the range of 0.3 and 0.9.

To explain the relation shown in equation (20) and (22), we apply 
a chemically inhomogeneous two-component model presented by \citet{liubarlow2000}.
In their hypothesis, the nebula consists of two components,
H-deficient clumps and diffuse material with `normal' abundances ($\sim$solar).
These H-deficient clumps are expected to have a low electron temperature
due to enhanced cooling caused by high metallicity and pressure equilibrium
between the two components, so that only recombination lines are emitted.
As a result, ORLs yield overestimated heavy element abundances.
In a given PN, the contribution of the two components
to the Balmer temperature and abundance determinations depends on
their densities, i.e., the contribution becomes weaker as density declines. 
If the clumps have neutral cores, as found in Abell~30 
\citep{ercolano} and NGC~7293 \citep{dyson},
they are expected to be quite stable during the evolution of the PNe.
As a result, the relative contribution of the
low-density and high-temperature material to the average temperature and
abundance determinations decrease as nebulae expand. Consequently, 
the discrepancies in temperature and abundance determinations become
larger as the nebulae expend.

\subsection{Are He/H ratios generally overestimated?}

A long-standing problem in abundance determinations for PNe has been that
the heavy element abundances derived from
ORLs are systematically higher than those derived from CELs. The ratio of
abundances derived from the two types of emission lines
 is found to vary from target to target.
\citet{tsamis} analysed a PN sample and found the ORL/CEL discrepancy factors
for these heavy element abundances span a range from about 2 to as large as
20. For the most extreme known PNe Hf~2-2, the ORL/CEL abundance ratio reaches
a record value of 84 \citep{liu209}. Another problem in nebular astrophysics
is that $T_{\rm e}$([\ion{O}{3}]$_{\rm na}$) are systematically higher than
$T_{\rm e}({\rm Bal})$, as we have discussed in \S~4.2. The two problems are 
found to be related. A positive correlation between the ratio of O$^{2+}$/H$^+$ derived
from \ion{O}{2} ORLs and from [\ion{O}{3}] CELs
and the temperature discrepancy [$\Delta T$=$T_{\rm e}$([\ion{O}{3}]$_{\rm na}$)-$T_{\rm e}({\rm Bal})$] has been shown by 
\citet{liu2001}.

The explanations for the dichotomy in abundance and temperature determinations
have focussed on temperature and density 
variations and chemical inhomogeneities. However, it has been shown that 
the temperature and density variations alone can be ruled out by
current observations \citep{liua,liu209}. 
At present, the most plausible explanation is that there exist
some cold H-deficient clumps in nebula \citep{liubarlow2000}
where only recombination lines are emitted.

To fit the observed pattern of NGC~6153, \citet{liubarlow2000} have 
presented several empirical models incorporating H-deficient clumps.
\citet{pequignot} also constructed photoionization
models of NGC~6153 and M~1-42, incorporating H-deficient clumps , 
which reproduce the observed integrated line 
ratio satisfactorily. In all the models, the H-deficient clumps contain only
about $1\%$ of the total mass, suggesting that the H-deficient
components contain only a tiny amount of nebular gas. Nevertheless, they
can seriously enhance the intensities of ORLs, causing overestimated ORL
abundances.
Therefore, the He/H abundance ratio could be overestimated owing to
the existence of the H-deficient clumps can lead to 
since all the observable strong \ion{He}{1} and \ion{He}{2} lines
are recombination lines. 
Some H-deficient components have been found in some PNe, such as Abell~30 and
Abell~78 \citep{jacoby}. \citet{wesson}
analysed optical spectra of the H-deficient knots in Abell~30 and found the
He/H abundance ratios of the knots are more than ten.

These PNe with extremely large abundance and temperature discrepancies, such
as NGC~6153, M~1-42 and Hf~2-2, have been found to have high He/H abundances.
\citet{peimbert1983} defined
the PNe having high He/H abundance ($\ge0.125$) as Type-I PNe.
According to our suggestion, these PNe with high He/H abundances might
result from contamination from H-deficient clumps. 
Quantitative analysis of overestimated He/H abundances in PNe is quite
difficult. For this purpose, as mentioned by \citet{pequignot}, one has
to rely on the measurement of \ion{He}{1} $\lambda$10830 line which is
strongly enhanced by collisional excitation from the 2s$^3$S metastable
level.

\begin{figure*}
\epsfig{file= 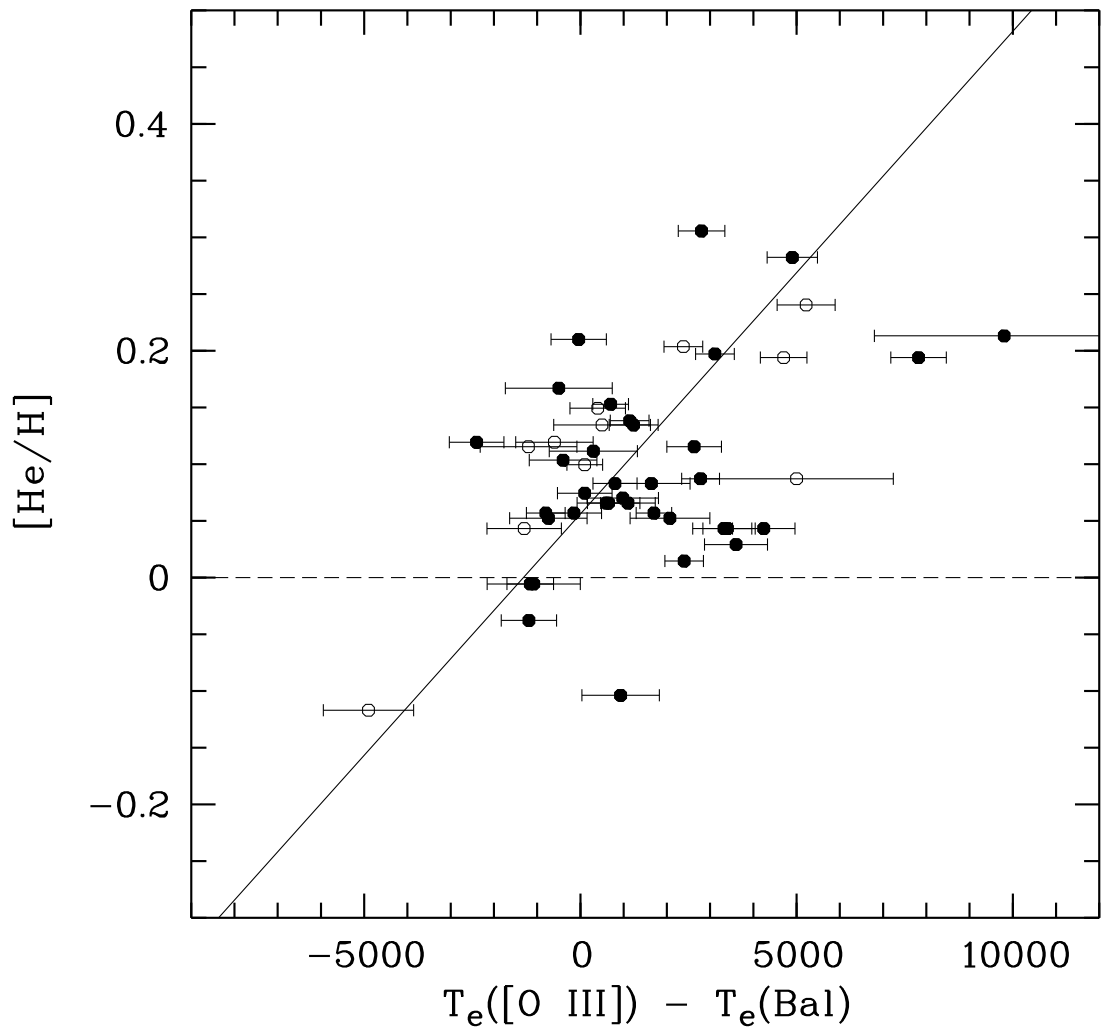,
height=14cm, bbllx=40, bblly=147, bburx=520, bbury=607, clip=, angle=0}
\caption{He/H versus $T_{\rm e}$([\ion{O}{3}])$ - T_{\rm e}$(Bal). 
The filled and open circles are the Galactic disc and bulge PNe, respectively. 
The solid line is a least-squares fit.
The dashed line is the solar He/H abundance. A
trend of increasing He/H abundance with increasing difference between
the temperature from [\ion{O}{3}] forbidden lines and  Balmer discontinuity
is seen, indicating that He/H ratios in PNe may be overestimated generally.
}
\label{hete}
\end{figure*}

If the He/H abundance is enhanced by H-deficient clumps, 
one expects that there is a
positive correlation between the He/H abundance and the ORL/CEL abundance ratio
of heavy element. For most PNe, however, heavy element ORLs are quite
weak and thus yield large uncertainties in the resultant abundance.
Instead, we can study the relation between the
He/H abundance and $\Delta T$ since the ORL/CEL abundance ratio
has been found to be positively correlated with $\Delta T$ by \citet{liu2001}.
Fig.~\ref{hete} plots He/H versus $\Delta T$. The measurement error of the
He/H abundance ratio is negligible. In the figure, filled and
open circles represent the Galactic disc and bulge PNe respectively and
the solid line is a linear fit represented by the relation:
\begin{equation}
{\rm He}/{\rm H}=(0.056\pm0.003)+(4.25\pm0.11)\times10^{-5}\Delta T,
\end{equation}
which yields a linear correlation coefficient of 0.50. Although there is considerable 
scatter, there
is a trend of increasing He/H abundance with increasing difference between
the temperature from [\ion{O}{3}] forbidden lines and Balmer discontinuity
for both the Galactic disc and bulge PNe. 
Therefore, it seems plausible that the He/H abundances in PNe have
been generally overestimated due to the existence of cold H-deficient clumps.

Fig.~\ref{hete} also shows that the most of PNe have 
enhancement of helium with respect to the Sun, which is often ascribed to
the second and third dredge-up. However, the normal model of dredge-up
cannot explain some extreme cases, such as He~2-111 which has a He/H
abundance of 0.219 \citep{kingsburgh}. For these cases, the more
likely reason is that pollution of the H-deficient clumps causes an
overestimate of the He/H abundance ratio. On the other hand, we cannot completely reject
the possibility that helium in nebulae has been partly enhanced by the 
products of nucleosynthesis of the central star. That may be the main cause
of the scatter in the derived He/H abundances. The presence of
hydrogen-deficient clumps may be further proof.

The ORL/CEL abundance discrepancies are also found in \ion{H}{2} regions
\citep{esteban,tsamisa}. And the existence of some high density clumps in 
\ion{H}{2} regions has been indicated by previous studies 
(e.g. O'Dell et al. 2003). Therefore, the H-deficient material might also
be present in \ion{H}{2} regions, leading to overestimated ORL abundances. 
If this is the case, the primordial helium
abundance, which is based on observations of some metal
poor extragalactic \ion{H}{2} regions, could have been overestimated.

\section{Temperatures and densities derived from the hydrogen recombination 
spectra near the Paschen jump}

\begin{figure*}
\epsfig{file=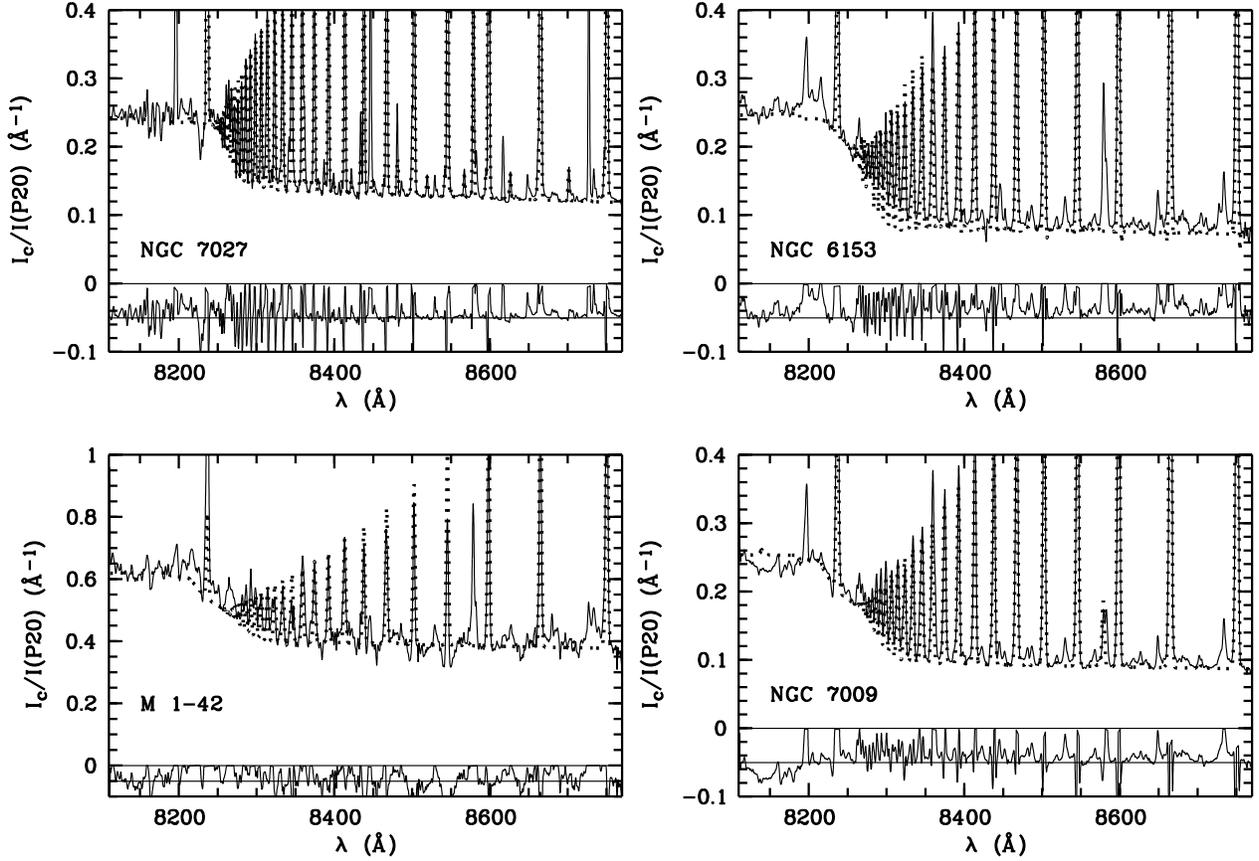,
height=11.5cm, bbllx=28, bblly=413, bburx=548, bbury=771, clip=, angle=0}
\caption{A comparison of 4 observed and theoretical spectra at 8250~\AA.
The top of each panel shows the
observed spectrum (solid line) which has been deredden using the extinction
constant listed in Table~\ref{tbl-1} and the synthetic spectrum (dot line).
The bottom shows residuals subtracted by 0.05. Note the
absorptions near 8200~\AA~~by atmospheric gases (H$_2$O and O$_2$)
and two \ion{Ca}{2} absorption features at 8500~\AA~~and
8546~\AA~~in the spectrum of M~1-42.}
\label{pash}
\end{figure*}

The hydrogen recombination spectrum near the Paschen jump can also
provides a diagnostic of electron temperature and density
[hereafter, $T_{\rm e}({\rm Pas})$ and $N_{\rm e}({\rm Pas})$]. Unfortunately,
in our sample, the spectra covering the Paschen jump are available for only 
four PNe, NGC~7027, NGC~6153, M~1-42 and NGC~7009. Using the same method 
described in Section~3, we match the theoretical with the observed spectra
at 8250~{\AA}. Given that the observed spectra near the
Paschen discontinuity may include a contribution from a dust emission tail,
we added a constant continuum for the match. The results are shown 
in Fig.~\ref{pash}. 
All the spectra are normalized to P~20 in order to decrease the uncertainty of reddening 
correction. Table~\ref{tbl-4} lists the derived temperatures and 
densities. 
One advantage of using the spectrum near the Paschen jump is that the resultant 
temperature and density are not affected by the continuum of the hot
central star and two-photon emission. However, the emission from local dust
may be stronger in this band.
In addition, the estimation of the continuum bluewards the Paschen jump is
 affected by the telluric absorption (H$_2$O and O$_2$). 
And the \ion{Ca}{2} absorption feature might also cause a error in fitting the continuum 
longward of the Paschen jump of some PNe, such as M~1-42.
As a result, the temperatures and densities derived from the hydrogen 
recombination spectra at the Paschen jump
are less reliable than those derived from those at
the Balmer jump.

Table~\ref{tbl-4} shows that the $N_{\rm e}({\rm Pas})$
are consistent with the $N_{\rm e}({\rm Bal})$
within the errors. However, the 
$T_{\rm e}({\rm Pas})$ are found to be 4000, 1000, 500 and 1200\,K 
lower than the $T_{\rm e}({\rm Bal})$ for NGC~7027, NGC~6153, M~1-42
and NGC~7009, respectively.
This may be due to systematic errors in the reddening correction
or flux calibration. As an alternative, it remains possible that 
the discrepancy is caused by temperature
variations. Further investigations including a 
larger data sample of PNe is needed to clarify the point.

\section{Summary}

In this work, we presented a method to derive nebular electron 
temperature and density
simultaneously by matching the theoretical with the observed hydrogen
recombination spectra. Using the method, we derived electron
temperatures and densities of
48 Galactic PNe. The temperatures and densities derived by this method
are compared with those derived from other indicators, enabling us to
study electron temperature and density variations in PNe. 
Our main findings are as follows.

\begin{list}{}{}
 \item[---] The densities derived from 
hydrogen recombination spectra are generally higher than those deduced from
forbidden line ratios, suggesting that condensations are 
generally present in nebulae.

\item[---] The temperatures 
derived from the nebular-auroral line ratios
of [\ion{O}{3}] are systematically higher than those derived from hydrogen 
recombination spectra, suggesting that temperature variations are 
generally present in nebulae. We
obtained the rms temperature fluctuation parameter $t^2=0.031$ as a 
representative value.

\item[---]The larger discrepancy in temperatures derived from 
[\ion{O}{3}] $(52\micron+88\micron)/\lambda4959$ line ratios and from 
hydrogen recombination spectra, 
however, cannot be attributed simply to the temperature variations.
The presence of clumps with $N_{\rm e}\sim10^5$~cm$^{-3}$ can account well
for the discrepancy. A comparison between electron temperatures derived
from [\ion{O}{3}] $(52\micron+88\micron)/\lambda4959$ line ratios and
those derived from hydrogen recombination spectra and
[\ion{O}{3}] $\lambda4959/\lambda4363$ line ratios suggests that the filling factor of the condensations
has a typical value of $7\times10^{-5}$.

\item[---]The discrepancies in temperatures derived from
[\ion{O}{3}] $\lambda4959/\lambda4363$ line ratios and from 
hydrogen recombination spectra are related to the evolution
of PNe. The ORL/CEL abundance ratio is found to have a power-law dependence on
electron density ($\sim N_{\rm e}^\alpha$, $\alpha=0.59\pm0.15$). We attribute
this to different evolutionary scenario of two components in PNe, i.e.
expanding diffuse nebular material and stable cold H-deficient 
clumps.
The contribution of the clumps to the 
temperature and abundance determinations increases as the nebulae expand.

\item[---]He/H abundance has a positive correlation with the temperature
discrepancy, suggesting that He/H abundances might be overestimated due to the
existence of H-deficient clumps. If this is the case, traditional up-dredge
theory should be modified. The primordial helium abundance determination
will be affected if there are significant condensations in \ion{H}{2} regions. 
Further investigations are needed.

\item[---] The densities derived from the spectrum near the Paschen jump are
in good
agreement with those derived form that near the Balmer jump. However, the
former yield generally lower temperatures than the latter. The reason
remain unknown.
\end{list}

We conclude that the existence of H-deficient clumps in PNe is the most
plausible explanation for the origin of temperature and
density variations and the ORL/CEL abundance discrepancies. However, it is a
challenge to clarify the origin and evolution of these clumps.
In further work, we will study the 
statistical relationships between the temperature and abundance
discrepancies and other physical conditions of PNe such as radio
flux, morphology, chemical abundances and the nature of the central star.

\acknowledgments

The authors are grateful to W. Wang for his help with part of the data 
reduction. We also thank J.-S Chen and S.-G. Luo for their help with
the preparation of this paper.
Constructive suggestions and comments by the referee S. M. Viegas
were appreciated.
This work was supported in part by Beijing Astrophysics Center
(BAC).

%\clearpage

%% Use the figure environment and \plotone or \plottwo to include 
%% figures and captions in your electronic submission.

\clearpage

\clearpage 

\clearpage

\clearpage

\clearpage

\clearpage

\clearpage

\clearpage

\clearpage

\clearpage
%%%%%%%%%%%%%%%%%%%%%%%%%%%%%%%%%%%%%%%%%%%%%%%%%%%%%%%%%%%%%%%%%%%%%%%%%%
\begin{deluxetable}{lcccccc}
%\tabletypesize{\scriptsize}
\tablecaption{ Object and parameters; line fluxes in 
units of $10^{-12}$~erg~cm$^{-2}$~s$^{-1}$.\label{tbl-1}}
\tablewidth{0pt}
\tablehead{
\colhead{Source} & \colhead{$c$}   & 
 \colhead{He$^+$/H$^+$} & 
\colhead{He$^{2+}$/H$^+$}&
\colhead{T$_{\rm e}$([\ion{O}{3}]$_{\rm na}$)}&
\colhead{T$_{\rm e}$([\ion{O}{3}]$_{\rm fn}$)} &
\colhead{$\sigma_o$}   \\
 & & & & (K) & (K) & (Km/s) \\
}
\startdata
Hu 1-2   &0.512   &0.046    &0.079  &$19500\pm 300  $  &--             &144\\
IC 3568  &0.261   &0.095    &0.001  &$11370\pm 200  $  &$12600\pm200 $ &132\\
NGC 7027 &1.380   &0.058    &0.041  &$12600\pm 180  $  &$15800\pm3000 $&82 \\
NGC 7009 &0.130   &0.093    &0.011  &$9980\pm  180  $  &$ 8350\pm150 $ &132\\
NGC 40   &0.700   &0.062    &0.000  &$10600\pm 200  $  &$10000\pm200 $ &132\\
NGC 6210 &0.055   &0.099    &0.001  &$9680\pm  200  $  &$10400\pm200$  &148 \\
NGC 6826 &0.064   &0.099    &0.000  &$9350\pm  200  $  &$10600\pm200$  &132 \\
NGC 6572 &0.400   &0.110    &0.000  &$10600\pm 200  $  &$17100\pm200$  &132 \\
NGC 6741 &1.100   &0.081    &0.031  &$12600\pm 200  $  &$19000\pm2000$ &132 \\
NGC 6790 &1.100   &0.081    &0.003  &$12840\pm 200  $  &--             &132 \\
NGC 6884 &1.050   &0.081    &0.016  &$10800\pm 200  $  &$21000\pm1000$ &132 \\
NGC 7662 &0.180   &0.061    &0.038  &$13300\pm 200  $  &$13000\pm200 $ &156 \\
NGC 6543 &0.100   &0.117    &0.000  &$7940\pm  200  $  &$8520\pm150 $  &140\\
M 2-24   &0.800   &0.099    &0.005  &$21000\pm 1500 $  &--             &156\\
Cn 2-1   &1.063   &0.104    &0.004  &$10400\pm 500  $  &--             &132\\
NGC 6720 &0.200   &0.094    &0.017  &$10630\pm 200  $  &$11600\pm200$  &132\\
NGC 6620 &0.760   &0.099    &0.021  &$10400\pm 400  $  &--             &132\\
NGC 6567 &0.670   &0.105    &0.007  &$11400\pm 400  $  &$13800\pm300$  &132 \\
H 1-35   &1.360   &0.115    &0.001  &$10500\pm 500  $  &--             &132\\
H 1-50   &0.700   &0.100    &0.011  &$11300\pm 500  $  &--             &132\\
He 2-118 &0.158   &0.065    &0.000  &$13400\pm 300  $  &--             &148\\
M 1-20   &1.200   &0.094    &0.000  &$10700\pm 500  $  &--             &123\\
M 3-21   &0.570   &0.100    &0.007  &$11100\pm 100  $  &--             &140\\
M 3-32   &0.579   &0.120    &0.013  &$9200 \pm 200  $  &--             &140\\
IC 1297  &0.220   &0.079    &0.042  &$10700\pm 100  $  &--             &132\\
IC 4634  &0.454   &0.097    &0.000  &$10200\pm 100  $  &$12500\pm300$  &140\\
IC 4776  &0.265   &0.088    &0.000  &$11000\pm 200  $  &--             &136\\
IC 4997  &0.438   &0.139    &0.000  &$20000\pm 1500 $  &--             &127\\
NGC 5873 &0.004   &0.063    &0.040  &$12800\pm 100  $  &--             &127\\
NGC 5882 &0.380   &0.113    &0.003  &--                &--             &127\\
NGC 6302 &1.460   &0.105    &0.067  &$18900\pm 200  $  &$15500\pm500 $ &132\\
NGC 6818 &0.400   &0.050    &0.051  &$12600\pm 200  $  &$15700\pm300 $ &148 \\
NGC 6153 &1.300   &0.123    &0.011  &$9110 \pm 200  $  &$ 8380\pm200 $ &148\\
M 2-36   &0.270   &0.133    &0.003  &$8380 \pm 200  $  &$ 8380\pm160 $ &148\\
M 1-42   &0.700   &0.139    &0.009  &$9220 \pm 300  $  &$ 6870\pm200 $ &140\\
NGC 6778 &0.890   &0.155    &0.008  &$9200 \pm 300  $ &--       &140\\
Hf 2-2   &0.200   &0.131    &0.002  &$8820 \pm 500  $ &--       &132\\
DdDm 1   &0.137    &0.060  &0.000   &$  12330\pm700$  & --      &99\\
Hu 2-1   &0.777    &0.097  &0.000   &$  9850 \pm300$  & --      &99\\
IC 2003  &0.347    &0.054  &0.049   &$  12640\pm400$  & --      &99\\
IC 5217  &0.501    &0.088  &0.008   &$  11260\pm400$  & --      &115\\
Me 2-2   &0.343    &0.138  &0.000   &$  10960\pm300$  & --      &90\\
NGC 6803 &0.869    &0.112  &0.004   &$  9730 \pm300$  & --      &107\\
NGC 6807 &0.642    &0.084  &0.000   &$  10920\pm500$  & --      &107\\
NGC 6833 &0.000    &0.078  &0.000   &$  12810\pm400$  &  --     &99\\
NGC 6879 &0.401    &0.091  &0.003   &$  10400\pm600$  &  --     &99\\
NGC 6891 &0.287    &0.094  &0.000   &$   9320\pm300$  &  --     &115\\
Vy 1-1   &0.421    &0.094  &0.000   &$  10240\pm600$  &  --     &99\\
\enddata                                        
%% Text for table notes should follow after the \enddata but before
%% the \end{deluxetable}. Make sure there is at $least one \tablenotemark
%% in the table for each \tablenotetext.

\end{deluxetable}

%% If you use the table environment, please indicate horizontal rules using
%% \tableline, not \hline.
%% Do not put multiple tabular environments within a single table.
%% The optional \label should appear inside the \caption command.

\clearpage
\begin{deluxetable}{lcc}
%\tabletypesize{\scriptsize}
\tablecaption{Electron temperatures and densities derived
from the spectra near the Balmer jump and forbidden line ratios.
\label{tbl-2}}
\tablewidth{0pt}
\tablehead{
\colhead{Source} & \colhead{T$_{\rm e}$(Bal)}   & 
\colhead{$\log$N$_{\rm e}$(Bal)}\\
  & (K) & (cm$^{-3}$) 
}
\startdata
Hu 1-2   &$20000\pm1200 $    &$3.9\pm 0.3$   \\
IC 3568  &$9300\pm900   $    &$3.8\pm 0.2$   \\
NGC 7027 &$12000\pm400  $    &$5.1\pm 0.1$   \\
NGC 7009 &$7200\pm400   $    &$3.8\pm 0.1$   \\
NGC 40   &$7000\pm700   $    &$3.2\pm 0.3$   \\
NGC 6210 &$8700\pm800   $    &$3.8\pm 0.1$   \\
NGC 6826 &$8700\pm700   $    &$3.4\pm 0.1$   \\
NGC 6572 &$10300\pm1000 $    &$4.4\pm 0.2$   \\
NGC 6741 &$15000\pm600  $    &$3.6\pm 0.2$   \\
NGC 6790 &$14000\pm500  $    &$4.5\pm 0.1$   \\
NGC 6884 &$11600\pm400  $    &$3.8\pm 0.1$   \\
NGC 7662 &$12200\pm 600 $    &$3.4\pm 0.3$   \\
NGC 6543 &$6800 \pm 400 $    &$3.8\pm 0.2$   \\
M 2-24   &$16000\pm 2000 $   &$7.0\pm 0.7$   \\
Cn 2-1   &$10800\pm 600  $   &$4.8\pm 0.2$   \\
NGC 6720 &$8000 \pm 600  $   &$2.9\pm 0.3$   \\
NGC 6620 &$10000\pm 500  $   &$3.7\pm 0.2$   \\
NGC 6567 &$12000\pm 800  $   &$4.0\pm 0.5$   \\
H 1-35   &$10000\pm 1000 $   &$4.8\pm 0.2$   \\
H 1-50   &$12500\pm 1000 $   &$4.1\pm 0.3$   \\
He 2-118 &$18300\pm 1000 $   &$5.0\pm 0.5$   \\
M 1-20   &$12000\pm 700  $   &$4.2\pm 0.3$   \\
M 3-21   &$11000\pm 400  $   &$4.2\pm 0.1$   \\
M 3-32   &$4500 \pm 500  $   &$3.6\pm 0.2$   \\
IC 1297  &$10000\pm 400  $   &$4.0\pm 0.2$   \\
IC 4634  &$8500 \pm 400  $   &$4.3\pm 0.2$   \\
IC 4776  &$8600 \pm 400 $    &$4.8\pm 0.2$   \\
IC 4997  &$10200\pm 500 $    &$6.0\pm 0.2$   \\
NGC 5873 &$12000\pm 500 $    &$3.9\pm 0.1$   \\
NGC 5882 &$6800 \pm 500 $    &$4.0\pm 0.2$   \\
NGC 6302 &$16100\pm 500 $    &$4.9\pm 0.1$   \\
NGC 6818 &$12500\pm 600 $    &$3.7\pm 0.1$   \\
NGC 6153 &$6000 \pm 400 $    &$3.8 \pm0.2$   \\
M 2-36   &$6000 \pm 400 $    &$3.8 \pm0.1$   \\
M 1-42   &$4000 \pm 600 $    &$3.7 \pm0.2$   \\
NGC 6778 &$4300\pm 500  $    &$3.6 \pm0.2$   \\
Hf 2-2   &$1000\pm400   $    &$2.6 \pm0.5$   \\
DdDm 1   &$  11400\pm800$    &$ 3.8\pm0.2$   \\
Hu 2-1   &$  10000\pm500$    &$ 4.3\pm0.2$   \\
IC 2003  &$  11000\pm800$    &$ 3.0\pm0.5$   \\
IC 5217  &$  12000\pm800$    &$ 3.0\pm0.5$   \\
Me 2-2   &$  11000\pm500$    &$ 4.2\pm0.2$   \\
NGC 6803 &$  8500\pm400$     &$ 3.6\pm0.4$   \\
NGC 6807 &$  12000\pm1000$   &$ 3.5\pm0.6$   \\
NGC 6833 &$  14000\pm500$    &$ 4.1\pm0.3$   \\
NGC 6879 &$  7000\pm400$     &$ 3.3\pm0.3$   \\
NGC 6891 &$ 6000\pm600$      &$ 3.6\pm0.6$   \\
Vy 1-1   &$ 6000\pm600$      &$ 3.0\pm0.5$   \\
\enddata                      

%% Text for table notes should follow after the \enddata but before
%% the \end{deluxetable}. Make sure there is at least one \tablenotemark
%% in the table for each \tablenotetext.

\end{deluxetable}

\clearpage
\begin{deluxetable}{lccccc}
%\tabletypesize{\scriptsize}
\tablecaption{The evaluation of $\mu^2\omega$.\label{tbl-3b}}
\tablewidth{0pt}
\tablehead{
\colhead{Source} & \colhead{$F(T_{\rm Bal})$}   & \colhead{$F(T_{[{\rm O}~{\rm III}]_{\rm fn}})$}
&
\colhead{$F(T_{[{\rm O}~{\rm III}]_{\rm na}})$}&
\colhead{$(\mu^2\omega)_{\rm H}$} &
\colhead{$(\mu^2\omega)_{\rm L}$}
}
\startdata
IC 3568     &    0.527    &  0.254   &   0.315   &   1.174   &   0.253  \\
NGC 7027    &    0.281    &  0.169   &   0.254   &   0.708   &   0.532  \\
NGC 7009    &    1.210    &  0.720   &   0.433   &   0.730   &   0.     \\
NGC 40      &    1.345    &  0.433   &   0.374   &   2.408   &   0.     \\
NGC 6210    &    0.641    &  0.392   &   0.469   &   0.681   &   0.208  \\
NGC 6826    &    0.641    &  0.374   &   0.520   &   0.768   &   0.415  \\
NGC 6572    &    0.401    &  0.151   &   0.374   &   1.857   &   1.639  \\
NGC 7662    &    0.271    &  0.238   &   0.228   &   0.145   &   0.     \\
NGC 6543    &    1.505    &  0.688   &   0.872   &   1.300   &   0.282  \\
NGC 6720    &    0.836    &  0.303   &   0.374   &   1.981   &   0.246  \\
IC 4634     &    0.688    &  0.258   &   0.411   &   1.867   &   0.634  \\
NGC 6818    &    0.258    &  0.171   &   0.254   &   0.543   &   0.516  \\
NGC 6153    &    2.543    &  0.714   &   0.254   &   2.987   &   0.     \\
M 2-36      &    2.543    &  2.056   &   0.714   &   0.249   &   0.     \\
M 1-42      &    2.543    &  1.422   &   0.543   &   0.850   &   0.     \\
Average     &             &          &           &   1.0     &   0.4    \\
\enddata                                                                
\end{deluxetable}
\clearpage

\begin{deluxetable}{lcc}
%\tabletypesize{\scriptsize}
\tablecaption{Electron temperatures and densities derived
from the spectra near the Paschen jump.\label{tbl-4}}
\tablewidth{0pt}
\tablehead{
\colhead{Source} & \colhead{T$_{\rm e}$(Pas)}   &
\colhead{$\log$N$_{\rm e}$(Pas)} \\
                 &    (K)     &  (cm$^{-3}$)}
\startdata
NGC 7027 &  $8000\pm1000$   &   $5.4\pm0.7$      \\
NGC 6153 &  $5000\pm700$    &   $3.8\pm0.5$      \\  
M 1-42   &  $3500\pm500$    &   $3.0\pm0.8$      \\
NGC 7009 &  $5800\pm900$    &   $3.5\pm0.7$      \\
\enddata               
\end{deluxetable}
%% Tables may also be prepared as separate files. See the accompanying
%% sample file table.tex for an example of an external table file.
%% To include an external file in your main document, use the \input
%% command. Uncomment the line below to include table.tex in this
%% sample file. (Note that you will need to comment out the \documentclass,
%% \begin{document}, and \end{document} commands from table.tex if you want
%% to include it in this document.)

%% \input{table}

%% The following command ends your manuscript. LaTeX will ignore any text
%% that appears after it.

\end{document}